\RequirePackage{fix-cm}
\documentclass[smallextended]{svjour3}       
\smartqed  
\usepackage{graphicx}
\usepackage{xcolor,soul,framed} 
\usepackage{multirow}
\usepackage{booktabs}
\colorlet{shadecolor}{yellow}
\usepackage[cmex10]{amsmath}
\usepackage{array}
\usepackage{mdwmath}
\usepackage{mdwtab}
\usepackage{eqparbox}
\usepackage{url}

\usepackage{tabu}
\usepackage[justification=centering]{caption}
\usepackage{subfigure}
\usepackage{graphicx}
\usepackage{bm}
\usepackage{amsfonts}
\begin{document}

\title{Revisiting Heterogeneous Defect Prediction: How Far Are We?
}

\author{Xiang~Chen \and
       Yanzhou~Mu \and
	  Chao~Ni \and
	  Zhanqi~Cui
}


\institute{Xiang~Chen \at
              School of Information Science and Technology, Nantong University, Nantong, China\\
              \email{xchencs@ntu.edu.cn}           
           \and
           Yanzhou~Mu \at
              College of Intelligence and Computing, Tianjin University, Tianjing, China\\
              \email{myz121796@gmail.com}
           \and
           Chao~Ni \at
           State Key Laboratory for Novel Software Technology, Nanjing University, Nanjing, China\\
           \email{jacknichao920209@gmail.com}
           \and
           Zhanqi~Cui \at
           Computer School, Beijing Information Science and Technology University, Beijing, China \\
           \email{czq@bistu.edu.cn}
}

\date{Received: date / Accepted: date}

\maketitle

\begin{abstract}
Cross-project defect prediction (CPDP) is applicable to the scenarios when the target projects are new projects or these projects only have a few training data. Most of the previous studies on CPDP tried to utilize the training data from other projects (i.e., the source projects) and then resorted to transfer learning to alleviate the data distribution difference between different projects.
However,  different metrics may be used by practitioners to measure the extracted program modules from different projects and performing CPDP across these projects with heterogeneous metrics  is more challenging. This issue is named as heterogeneous defect prediction (HDP).
Until now, researchers have proposed several novel HDP methods with promising performance.
Recently, unsupervised methods have been widely studies in CPDP and show the competitive performance. Moreover,
this kind of methods can be easily compared with HDP methods. To the best of our knowledge, whether HDP methods can  perform significantly better than these unsupervised methods has not yet been thoroughly investigated.
In this article, we perform a replication study to have a holistic look in this issue.
In particular, we compare state-of-the-art five HDP methods  with state-of-the-art five unsupervised methods.
Final results surprisingly show that these HDP methods do not perform significantly better than some of unsupervised methods (especially the simple unsupervised methods proposed by Zhou et al.) in terms of two non-effort-aware performance measures (i.e., $\mathit{F1}$ and $\mathit{AUC}$) and four effort-aware performance measures (i.e., $\mathit{ACC}$, $\mathit{P_{opt}}$, $\mathit{PMI@20\%}$ and $\mathit{IFA}$). Then, we perform diversity analysis on defective modules via McNemar's test and find the prediction diversity is  more obvious when the comparison
is performed between the HDP methods and the unsupervised
methods than the comparisons only between the HDP methods or between the unsupervised methods. This shows the HDP methods and the unsupervised methods are complementary  to each other in identifying defective modules to some extent. Finally, we investigate the feasibility of five HDP methods by considering
two satisfactory criteria recommended by previous CPDP studies and find the  satisfactory ratio of these
HDP methods  is still pessimistic.
The above empirical results implicate there is still a long way for heterogeneous  defect prediction to go. More effective HDP methods need to be designed and the unsupervised methods should be considered as baselines.

\end{abstract}

\section{Introduction}
\label{intro}
Software defect prediction (SDP)~\cite{wan2018perceptions,hall2011systematic,kamei2016defect} is an active and thriving area in the domain of software repository mining.
SDP methods can be used to optimize the  software testing resource allocation (e.g., designing more high-quality test cases or performing more rigorous code inspection) by identifying potential defective modules in advance.
Most of the previous SDP studies focus on within-project defect prediction (WPDP), which builds the SDP models and predicts defective modules within  the same project.
However, the difficulty of gathering SDP datasets (especially labeling program modules as defective or non-defective by using the SZZ approach~\cite{da2016framework}) is the main obstacle of applying SDP to the practical software quality assurance process of enterprises~\cite{Sliwerski:2005}.
In practice, the target projects for SDP are often new projects or these projects only have a few training data in most cases. In previous studies~\cite{Hosseini:TSE2017}, researchers mainly consider to utilize the training data from other projects (i.e., the source projects) and utilize transfer learning~\cite{pan:TKDE2010} to alleviate the data distribution difference between different projects.
This problem is called as cross-project defect prediction (CPDP).

Most of the previous studies on CPDP assume that both the source project and the target project use the same metrics to measure the extracted program modules (i.e., homogeneous data).
In reality, practitioners  may use different metrics to measure the extracted program modules in their projects. The reasons can be summarized as follows.
First, some metrics are designed based on a specific programming language, therefore these metrics cannot be used to measure program modules written by other programming languages. Second, some metrics can be only supported by commercial tools (such as Understand tool\footnote{https://scitools.com/}) and the cost of buying these commercial tools  cannot be afforded for small-scale start-up enterprises~\cite{nam2017heterogeneous}.
Compared to CPDP across the projects with homogeneous metrics,  performing CPDP across the projects with heterogeneous metrics is more challenging and this problem is named as heterogeneous defect prediction (HDP)~\cite{nam2017heterogeneous}.

Until now, researchers have proposed several novel methods for solving the HDP problem. For example, Nam et al.~\cite{nam2015heterogeneous,nam2017heterogeneous} proposed a HDP method based on metric selection and metric mapping. Li et al.~\cite{li2018cost,li2017heterogeneous,LiASEJ2019} proposed a set of novel HDP methods based on kernel correlation
alignment and ensemble learning.
Moreover, these researchers have even shared code of these methods for the convenience of replicating their studies.
Recently, some studies~\cite{zhou2018far,Zhang:ICSE2016,Nam:ASE2015,ni2019empirical} have showed the competitiveness of unsupervised methods for CPDP and this kind of methods can be easily compared with HDP methods. However, to the best of our knowledge, this issue has not been thoroughly investigated in previous HDP studies~\cite{nam2015heterogeneous,nam2017heterogeneous,li2018cost,li2017heterogeneous,LiASEJ2019}.
In our empirical studies, we choose five state-of-the-art HDP methods and five state-of-the-art unsupervised methods.
In particular, the chosen five HDP methods include the method based on metric selection and matching proposed by Nam et al.~\cite{nam2015heterogeneous,nam2017heterogeneous}, the three methods based on metric representation proposed by Li et al.~\cite{li2018cost,li2017heterogeneous,LiASEJ2019} and the method based on distribution characteristics proposed by He et al.~\cite{Hearxiv14}.
The chosen five unsupervised methods include the method CLA and CLAMI proposed by Nam and Kim~\cite{Nam:ASE2015}, the connection based method proposed by Zhang et al.~\cite{Zhang:ICSE2016}, ManualUp and ManualDown methods  proposed by Zhou et al.~\cite{zhou2018far} and their variants.
Then we choose five groups of datasets (i.e., AEEEM, ReLink, PROMISE, NASA and SOFTLAB), which include 34 different projects in total. These datasets have been widely used in previous HDP studies~\cite{nam2015heterogeneous,nam2017heterogeneous,li2018cost,li2017heterogeneous,LiASEJ2019}.

To systematically investigate this issue, we aim to answer the following four research questions.

\noindent \textbf{RQ1: Do these HDP methods perform significantly better than existing unsupervised methods in terms of non-effort-aware performance measures (NPMs)?}

For RQ1, we mainly consider two NPMs (i.e., $\mathit{F1}$ and $\mathit{AUC}$) and use Scott-Knott test~\cite{jelihovschi2014scottknott} to rank all the HDP methods and the unsupervised methods. Final results show HDP methods cannot perform significantly better than all the unsupervised methods in terms of these two NPMs.

\noindent \textbf{RQ2: Do these  HDP methods perform significantly better than existing unsupervised methods in terms of effort-aware performance measures (EPMs)?}

For RQ2, we mainly consider four EPMs. In particular, The first two EPMs are
$\mathit{P_{opt}}$ and  $\mathit{ACC}$, which were previously used for just-in-time defect prediction~\cite{kamei2013large,Yang:FSE2016,chen2018multi}.
The latter two EPMs are $\mathit{PMI@20\%}$ and $\mathit{IFA}$, which were proposed by Huang et al.~\cite{Huang:ICSME2017,huang2018revisiting}. We also use Scott-Knott test~\cite{jelihovschi2014scottknott} to rank all the HDP methods and the unsupervised methods.
Final results also show HDP methods cannot perform significantly better than some of unsupervised methods in terms of these four EPMs.

\noindent \textbf{RQ3: How about the diversity analysis results on identifying defective modules between the HDP methods and existing unsupervised methods?}

For RQ3, we use McNemar's test~\cite{dietterich1998approximate} to perform diversity analysis on identifying defective modules. Final results show the prediction diversity is more obvious when the comparison is performed between the HDP methods and the unsupervised methods than the comparison  only performed between the HDP methods or between the unsupervised methods. This shows these two kinds of methods are complementary to each other in identifying defective modules to some extent. Moreover, we also find that there exists a certain number of defective modules, which cannot be correctly identified by either the HDP method or the unsupervised methods. These findings implicate
more effective HDP methods need to be designed in the future.

\noindent  \textbf{RQ4:  Whether the HDP methods can achieve satisfactory performance?}

For RQ4, we mainly consider the satisfactory criteria suggested from previous studies for CPDP~\cite{zimmermann:FSE2009,he2012investigation}. Final results show that the satisfactory ratio of these
HDP methods on different groups of datasets is still pessimistic.


The main contributions of our article can be summarized as follows.

\begin{itemize}
\item To the best of our knowledge, we are the first to design and conduct large-scale empirical studies to systematically compare the existing five HDP methods with state-of-the-art five unsupervised methods on five groups of datasets (34 different projects in total) under the same experimental
setup.

\item We make a comprehensive comparison between the HDP methods and the unsupervised methods  from three different perspectives: non-effort-aware performance measures, effort-aware performance measures and diversity analysis on identifying defective modules. Moreover, we also analyze  whether the HDP methods can achieve satisfactory performance based on two satisfactory criteria proposed in previous CPDP studies. Final results implicate there is still a long way for
HDP to go.

\end{itemize}

The rest of this article is organized as follows.
Section~\ref{chap2} analyzes the background of software defect prediction and related work for cross-project defect prediction.
Section~\ref{chap3} introduces the details and experimental setting of our chosen heterogeneous defect prediction methods and the unsupervised methods.
Section~\ref{chap4} shows the experimental setup, including research questions, experimental subjects and performance measures.
Section~\ref{chap5} performs result analysis and threats to validity analysis. Section~\ref{chap6} concludes this article and points out some potential future work.

\section{Background and Related Work}
\label{chap2}

This section first introduces the background of software defect prediction and its main process of the model construction phase and the model application phase, then it summarizes the related work for cross-project defect prediction, including heterogeneous defect prediction studied in our study.

\subsection{Background of Software Defect Prediction}
\label{chap2.1}

Software defects may be introduced in the source code during different phases of software development process. The root cause of defects may come from misunderstanding  the clients' requirements, irrational development process, or the lack of developers' experience in programming languages or domain knowledge. Hidden defects in the projects will produce unexpected results and even result in huge economic loss for enterprises in the worst cases after these projects are deployed.

Since available software quality assurance (SQA) resources allocated for software testing and code inspection are usually limited, effective methods are required by SQA groups to identify potential defective software modules as early as possible, which help to optimize the allocation of these limited SQA resources.
Software defect prediction is one of such feasible and effective methods.
The prediction target of SDP is often set to defect-proness or defect number/density for the new program modules according to the real application needs.
If the prediction target is set to defect-proness,  the process of SDP can be summarized as follows:
software modules are firstly automatically extracted from software historical repositories, such as version control systems (e.g., SVN, CVS, GIT) and bug tracking systems (e.g., Bugzilla, Jira), since the version control systems contain the source codes
and commit messages, while the bug tracking systems
include bug reports.
Secondly, software metrics (i.e., features) are designed and then used to measure these extracted program modules and the type (i.e., defective or non-defective) of the modules are  labeled via the SZZ method~\cite{da2016framework} by analyzing bug reports and commit messages. Metrics are often designed in the manual way based on the analysis of the code complexity, the characteristics of development process or the developers' experience~\cite{radjenovic2013software}. Thirdly, SDP models  are trained by using machine learning methods based on the gathered SDP datasets. Finally, the constructed SDP models can be utilized to predict defect-proneness (i.e., defective or non-defective) of new software program modules in the target project.

\subsection{Related Work for Cross-project Defect Prediction}
\label{chap2.2}

Most of the previous studies focused on within-project defect prediction (WPDP).
These studies build the prediction models and predict defects for new program modules within the same project.
In real software development, projects needing defect prediction may be  new projects or  have less training data. That  means these projects do not have sufficient historical data to construct high-quality prediction models.
A simple and straight solution is to directly use the training data gathered from other projects to construct the models.
However, application domain, utilized programming languages, development process, developers' experience of different projects may be not the same.
These difference will result in the non-negligible data distribution difference in most cases.

Researchers conducted a set of empirical studies to investigate the feasibility of CPDP by considering real-world software projects.
Zimmermann et al.~\cite{zimmermann:FSE2009}  analyzed 12 real-world projects from open-source communities and Microsoft corporation. After running 622 cross-project predictions, they found only 3.4\% (i.e., 21) can achieve satisfactory performance.
Later, He et al.~\cite{he2012investigation} analyzed another 10 open-source projects. After running 160,586 cross-project predictions, they found only 0.32\% to 4.67\% when consider different classifiers can achieve satisfactory performance.
Moreover, Kamei et al.~\cite{fukushima2014empirical,kamei2016studying} still found the unsatisfactory performance of CPDP in the context of just-in-time (change-level) software defect prediction~\cite{kamei2013large}.
Rahman et al.~\cite{rahman2012recalling} investigated the feasibility of CPDP in terms of effort-aware performance measures (i.e., taking the limitation of available SQA resources into consideration). They surprisingly found that the performance of CPDP is no worse than WPDP and  significantly better than the random method.
Turhan~\cite{turhan2012dataset} summarized the reasons of poor performance of CPDP  via dataset shift concept.
He classified different forms of dataset shift into 6 categories, such as covariate shift,
prior probability shift.
His analysis forms the theoretical support for the follow-up studies on CPDP.

Until now, researchers have proposed a number of  CPDP methods to reduce data distribution difference between the source  and  target projects~\cite{Hosseini:TSE2017,herbold:TSE2017}. In this subsection, we simply classify existing CPDP methods into 4 categories:
supervised homogeneous CPDP methods, supervised heterogeneous CPDP methods (i.e., HDP methods), semi-supervised CPDP methods and unsupervised CPDP methods. Then we introduce the related work for each category.

\subsubsection{Supervised Homogeneous CPDP Methods}

Most of CPDP studies fall into this category. In this category, researchers assumed that the source project and the target project use the same metrics to measure the extracted modules. Moreover, these methods only use the data in the source projects to construct the models.
(1)
Some  methods focus  on metric value transformation, such as logarithmic transformation,  min-max normalization, z-score normalization, Box-Cox transformation~\cite{camargo:ESEM2009,nam:ICSE2013,zhang:EMSE2017}. The objective of some metric value transformation methods is to make the transformed values of the metric satisfy the normal distribution assumption or make value distribution more symmetrical.
(2)
Some methods focus on selecting relevant source projects, which are  similar to the target project, from candidate source projects.
He et al.~\cite{he2012investigation} utilized distribution characteristics of metrics to choose relevant projects.
Krishna et al.~\cite{krishna2016too,krishna2018bellwethers} considered the usage of bellwether. Among candidate source projects, the bellwether is the project, which can achieve better performance than all the other projects.
Liu et al.~\cite{liu2018two} proposed a source project estimator (SPE) method. The SPE method can choose the projects from the candidate source projects with highest distribution similarity.
(3)
Some methods focus on instance selection or instance weight setting for the  modules in the source project.
Menzies et al.~\cite{menzies2011local,menzies:TSE2013}, Bettenburg et al.~\cite{bettenburg2012think,bettenburg2015towards} and Herbold et al.~\cite{herbold2017global} considered local models. These local models can cluster the available training data into homogeneous regions and then train the models by classifiers for each region.
Turhan et al.~\cite{turhan:EMSE2009} proposed Burak filter. Burak filter can select most similar modules from the source project via $k$-nearest neighbor.
Peters et al.~\cite{peters:MSR2013} later proposed Peters filter by analyzing the structure of the target project, while Burak filter mainly analyzed the structure of the target project.
Herbold~\cite{herbold2013training} utilized EM clustering and $k$-nearest neighbor to perform instance selection from the source project.
Hosseini et al.~\cite{hosseini2016search,hosseini2018benchmark} used genetic algorithm to perform instance selection.
Bin et al.~\cite{bin2017training} investigated more strategies for instance selection and surprisingly found  performing instance selection is not necessary.
Xu et al.~\cite{Xu:2018ICPC} proposed dissimilarity-based sparse feature selection method.
Ma et al.~\cite{ma:IST2012} proposed  Transfer Naive Bayes (TNB) method. Different from instance selection methods, TNB method can assign weights for the instances in the source project.
Poon et al.~\cite{poon2017cross}
proposed a credibility theory based Naive Bayes classifier to design a novel re-weighting mechanism.
(4)
Some methods focus on feature mapping and  feature selection~\cite{ni2019empirical,liu2015empirical}.
Nam et al.~\cite{nam:ICSE2013} applied a state-of-the-art transfer learning method (i.e., TCA~\cite{pan2011domain}) to make feature distribution
similar and then proposed TCA+, which can automatically select a suitable normalization method when given the specific source and target projects.
He et al.~\cite{He:IST2015} investigated the feasibility of the CPDP models constructed by a simplified feature subset.
Ni et al.~\cite{ni2017fesch,Ni:JCST2017} proposed a cluster based feature selection method FeSCH. In the
feature clustering phase, it first clusters features via a density-based clustering method. Then in the feature selection phase, it selects
features from each cluster by a ranking strategy.
(5) Some studies resort to  advanced machine learning methods. For example,
Panichella et al.~\cite{panichella:SANER2014} proposed a combined method CODEP based on ensemble learning. Then Zhang et al.~\cite{zhang2015empirical}  considered other 7 ensemble learning methods.
Since most of defects exist in a few modules for most of  projects under testing, there exists a certain class imbalanced problem in most of the gathered SDP data sets.
Ryu et al.~\cite{ryu2016value}\cite{ryu2015hybrid}\cite{ryu2017transfer} designed CPDP methods by considering the class imbalanced problem.
Limsettho et al.~\cite{limsettho2018cross} proposed a method CDE-SMOTE by using class distribution estimation and SMOTE (synthetic minority oversampling technique)~\cite{Chawla2002}.
Jing et al.~\cite{jing2017improved} considered subclass discriminant analysis (SDA) method.
Wang et al.~\cite{wang2018deep} resorted to deep learning. They utilized deep belief network (DBN) to automatically learn semantic features from token vectors extracted from ASTs (abstract syntax trees) of program modules.
Both Canfora et al.~\cite{canfora2013multi,canfora2015defect} and Ryu et al.~\cite{ryu2016effective}  used multi-objective optimization to construct CPDP models.
Wang et al.~\cite{Wang:apsec2018} proposed a top-$k$ learning to rank method, which can help to assign higher rank to modules with higher severity.

\subsubsection{Semi-supervised CPDP Methods}

Existing CPDP methods in this category assume that there are a  few   program modules in the target project are labeled and these modules can be utilized with the modules in the source project to construct models.
Turhan et al.~\cite{turhan2013empirical} investigated the feasibility of using mixed project data (i.e., modules from the source project and the target project).
Ryu et al.~\cite{ryu2017transfer} proposed a transfer cost-sensitive boosting method. This method
considers both knowledge transfer and class imbalanced learning when given a small
number of labeled modules in the target project.
Xia et al.~\cite{xia:TSE2016}
proposed a hybrid model construction method
HYDRA. HYDRA method includes the genetic algorithm phase and the ensemble learning
phase.
Chen et al.~\cite{chen2015negative} proposed
DTB (double transfer boosting) method. DTB method first used a data gravitation method to reshape the whole distribution of data in the source project to fit the data in the target project. Then DTB method used the transfer boosting method  to utilize a few labeled modules in the target project to identify and eliminate irrelevant instances in the   source project.
Zhang et al.~\cite{zhang2017label} utilized graph based semi-supervised learning method.
Wu et al.~\cite{wu2018cross} utilized a semi-supervised dictionary learning method and then proposed a cost-sensitive
kernelized semi-supervised dictionary learning method.

\subsubsection{Unsupervised CPDP Methods}

The CPDP methods in this category attempt to perform defect prediction on the modules in the target project immediately and these methods do not need any label information.
The assumption of these unsupervised methods is that the metric values of defective modules have the tendency to be higher than the metric values of non-defective modules~\cite{Nam:ASE2015,Zhang:ICSE2016}.
Nam and Kim~\cite{Nam:ASE2015} proposed CLA and CLAMI methods. The key idea of these two methods was to label an unlabeled
dataset by using the magnitude of metric values.
Zhang et al.~\cite{Zhang:ICSE2016} proposed a connectivity-based  method via spectral clustering.
Recently, Zhou et al.~\cite{zhou2018far} found that simple unsupervised methods (i.e., ManualDown and ManualUp) have better prediction performance than complex CPDP methods. Their findings indicated that previous studies on CPDP seem to make a simple problem complex.

\subsubsection{Supervised Heterogeneous CPDP Methods}

In this article, we mainly focus on supervised heterogeneous CPDP methods (i.e., HDP methods).
Researchers found that different practitioners may use different metrics to measure the extracted modules and this results in the hypothesis of the methods in traditional  supervised homogeneous  CPDP methods invalid.
It is not hard to find that this problem is more challenging than supervised homogeneous  CPDP problem.
A simple solution is to use the common metrics used by both the source project and the target project. However, the number of the common metrics may be very small, therefore some informative metrics for constructing high-quality models may not exist in the common metrics.
Moreover, finding other projects with similar metrics is also a challenging task.
He et al.~\cite{Hearxiv14} proposed a distribution characteristic based HDP method.
This method considers 16 distribution characteristics, such as mode, median.
Nam et al.~\cite{nam2015heterogeneous,nam2017heterogeneous} proposed HDP method to perform defect prediction across projects with heterogeneous metrics. This method includes the metric selection phase and the metric matching phase. Then Yu et al.~\cite{yu2017feature}
presented a feature matching method to convert
the heterogeneous features into the matched features, and
 presented a feature transfer method to transfer the matched
features from the source project to the target project.
Jing et al.~\cite{Jing2015Heterogeneous} proposed UMR (unified metric representation) for the data of the source project and the target project, then they used CCA (canonical correlation analysis) to make the data distribution similar.
Li et al.~\cite{li2018cost} proposed a new  cost-sensitive
transfer kernel canonical correlation analysis method. This method can not only make the data distribution more similar, but also utilize the different misclassification costs for defective and non-defective modules to alleviate the class imbalanced problem.  Meanwhile, they~\cite{li2017heterogeneous} proposed
a novel EMKCA (ensemble multiple
kernel correlation alignment) based method. Recently, they~\cite{LiASEJ2019} proposed a two-stage ensemble learning (TSEL) based
approach, which includes ensemble multi-kernel
domain adaptation  phase and ensemble data sampling
 phase.
 Since there exist multiple candidate source projects,
 Li et al.~\cite{Li2017On} proposed a multi-source selection based manifold discriminant alignment (MSMDA) based approach. MSMDA approach can incrementally select distribution-similar
source projects for a given
target project. Moreover, they designed a sparse representation based double
obfuscation algorithm to protect the privacy of dataset owners.
Li et al.~\cite{Li2018Heterogeneous}  proposed cost-sensitive label and structure-consistent
unilateral projection (CLSUP) based approach. CLSUP approach exploited the mixed project data (i.e., combine the heterogeneous source and target project data) and aimed to
transform the source data to the target subspace, where the data distributions of source and target projects become similar and the structure
of source data can be maintained. Then it used different misclassification costs for defective and non¨Cdefective modules in the
domain adaptation stage to alleviate class imbalanced problem.

Recent studies~\cite{Nam:ASE2015,Zhang:ICSE2016,zhou2018far} have showed the competitiveness
of unsupervised methods for CPDP. However, whether HDP can perform significantly better than these unsupervised methods has not been thoroughly investigated.  In this article, we choose  state-of-the-art five HDP methods and five unsupervised methods. The comparisons are systematically conducted from three different perspectives: non-effort-aware performance measures, effort-aware performance measures and diversity analysis on identifying defective modules. Final results show that studies on the heterogeneous defect prediction still have a long way  to
go and simple unsupervised methods (such as the unsupervised method proposed by Zhou et al.~\cite{zhou2018far}) should be considered as baselines when researchers evaluate their proposed new HDP methods.

\section{Methods}
\label{chap3}

This section first shows  the details of the  HDP methods considered by our study.
Then this section introduces the details of the unsupervised CPDP methods considered by our study.
Moreover, this section also introduces the experimental setup for these considered methods.

\subsection{HDP Methods}
\label{chap3.1}

Based on the related work analysis, researchers have proposed several novel HDP methods.
In our study, we mainly investigate the following five HDP methods,
since these  HDP methods have been published in refereed conferences or journals in software engineering research domain (such as TSE, ASE, ICSME) and these methods have been proposed in recent 5 years.
Some of HDP methods analyzed in related work are not considered in our study and the reasons can be summarized as follows. The method MSMDA~\cite{Li2017On} mainly focus on candidate source project selection and privacy protection of dataset owner. The method CLSUP~\cite{Li2018Heterogeneous} mainly focus on mixed data (i.e., combine the heterogeneous source and target project data). These two methods do not have the same experimental setup in our study, since our study only concerns the HDP combination with only a source project and a target project.
In this subsection, we will introduce the details of these HDP methods and experimental setup in our empirical studies.

\noindent \paragraph{\textbf{HDP1.} }
This method is proposed by Nam et al.~\cite{nam2015heterogeneous,nam2017heterogeneous}.
It first uses metric selection to the source project to identify and remove redundant and irrelevant features.
Later, it  measures the similarity of each source and target metric pair. Then it uses the cutoff threshold to remove poorly matched metrics.
After the metric mapping phase, it uses the maximum weighted bipartite matching technique to select a group of matched metrics.
Finally, it uses a classifier to
 build a prediction model using a source dataset with
selected and matched metrics. Then, this model can be used to predict defects on a target dataset
with metrics matched to selected source metrics.

In our study, the experimental setup for HDP1 is set as follows based on the suggestions by Nam et al.~\cite{nam2015heterogeneous,nam2017heterogeneous}. We use gain ratio for metric selection and
 select top 15\%  features.
Then we choose Kolmogorov-Smirnov test to measure the similarity with the cutoff threshold of 0.05 as the maximum weighted bipartite matching technique to select the best suitable group of matched metrics.
Finally, we use Logistic regression as our classifier.

\noindent \paragraph{\textbf{HDP2.} }
This is ensemble multiple kernel correlation alignment (EMKCA) based method and it is proposed by Li et al.~\cite{li2017heterogeneous}.
It first maps the data in the source project and the data in the target project into high dimensional kernel space through multiple kernel learning. Therefore, defective modules and non-defective modules in the high dimensional kernel space can be better separated.
Later, it utilizes a kernel correlation alignment method to make the data distribution of the source and target projects similar in the kernel space.
Finally, it integrates multiple kernel classifiers via ensemble learning to alleviate the class imbalanced problem.

In our study, the experimental setup for HDP2 is set as follows based on the suggestions by  Li et al.~\cite{li2017heterogeneous}.
In particular,
we use 10 base kernels $k(x_{i},x_{j})$, which include 9 Gaussian kernels $e^{-||x_i-x_j||/2\sigma^2}$ with different $\sigma$ in $\{2^{-4}, 2^{-3}, 2^{-2}, \cdots, 2^3, 2^4\}$ and a linear kernel on all the metrics.
Moreover, we set the value of the parameter $r$ in ICD (Incomplete Cholesky decomposition)~\cite{vaerenbergh2010kernel} to 60. Here ICD is a matrix decomposition method used to reduce kernel matrix. Finally  we use Logistic regression as our classifier.

\noindent \paragraph{\textbf{HDP3.} }
HDP3 is cost-sensitive transfer kernel canonical correlation analysis  (CTKCCA) based method and it is proposed by Li et al.~\cite{li2018cost}.  In particular, it first employ z-score normalization to preprocess data. Then it uses transfer kernel canonical correlation analysis for deriving the nonlinear feature space, where the learned features have more favorable separability. Finally it   utilizes cost-sensitive
learning technique, which considers the different misclassification costs for defective modules and
non-defective modules. Since the misclassification cost of predicting defective modules as non-defective modules is much higher than the misclassification cost of predicting non-defective modules as defective modules.

In our study, the experimental setup for HDP3 is set as follows based on the suggestions by  Li et al.~\cite{li2018cost}.
In particular,
we first use the Gaussian kernel function $k(x_{i},x_{j})=e^{-||x_i-x_j||/2\sigma^2}$ and set the kernel parameter $\sigma$ to the inverse of mean of square distance for the corresponding source project and target project respectively. Second
the misclassification
costs  in CTKCCA are adaptively set from 1 to $nondefective/defective$, where $nondefective$ denotes the number of non-defective modules and $defective$ denotes the number of defective modules in the corresponding source project.
Later, we   set the value of the parameter $r$ in ICD to 70.
 Finally  we use Logistic regression as our classifier.

\noindent \paragraph{\textbf{HDP4.}}
HDP4 is a two-stage ensemble learning (TSEL) based approach and it is proposed by Li et al.~\cite{LiASEJ2019}.
In particular,
this approach contains two phases: ensemble multi-kernel domain adaptation (EMDA) phase and
ensemble data sampling (EDS) phase. In the EMDA phase, it proposes an ensemble
multiple kernel correlation alignment (EMKCA) predictor, which combines the
advantage of multiple kernel learning and domain adaptation techniques. In the EDS
phase, it employs resample with replacement  technique to learn multiple different
EMKCA predictors and use average ensemble to combine them together.
At the end of EMDA and EDS phases, it has an ensemble
of defect predictors and it can be used to predict defective modules in the target project.

In our study, the experimental setup for HDP4 is set as follows based on the suggestions by  Li et al.~\cite{LiASEJ2019}. In particular,
there are three parameters in HDP4. The first  parameter $N$ is used to choose the number of EMKCA predictors and the value of this parameter is set to 10. The second parameter $M$ is used to decide the number of base kernels and the value of this parameter is set to 10 (i.e., these base kernels are 9 Gaussian kernels $e^{-||x_i-x_j||/2\sigma^2}$ with different $\sigma$ in $\{2^{-4}, 2^{-3}, 2^{-2}, \cdots, 2^3, 2^4\}$ and a linear kernel on all the metrics). The third parameter  $r$ is used in ICD and the value of this parameter is set to 60. Later HDP4 uses logistic regression classifier to construct prediction models.

\noindent \paragraph{\textbf{HDP5.} }
HDP5 is distribution characteristic based method and it is proposed by  He et al.~\cite{Hearxiv14}.
This method considers 16 distribution characteristics, including mode, median, mean, harmonic mean, minimum, maximum, range, variation
ratio, interquartile range, variance, standard deviation, coefficient of variation,
skewness and kurtosis. Then it employs these
distributional characteristic vectors of each program module as new metrics to build HDP models via logistic regression classifier
and predicts defective  modules in target project.

\subsection{Unsupervised  Methods}

In our article, we mainly consider five state-of-the-art unsupervised methods.
These  unsupervised methods have also been published in refereed conferences or journals in software engineering research domain (such as TOSEM, ASE, ICSE) and these methods have also been proposed in recent 5 years.

\noindent \paragraph{\textbf{UDP1 and UDP2.} }  Nam and Kim~\cite{Nam:ASE2015} proposed CLA (UDP1) and CLAMI (UDP2) methods.
The key idea of these two unsupervised methods was to label unlabeled dataset by using the magnitude of metric values.
Hence, CLA
and CLAMI have the advantages of automated manner and no manual efforts are required. The first two phases of these
two methods were (1) clustering unlabeled program modules, and (2) labeling these modules in clusters. CLA only consisted
of these two phases. CLAMI had two additional phases to generate the training dataset. (3) metric selection and (4)
instance selection, since these two phases can be used to further improve the quality of the datasets. Then CLAMI used Logistic
regression classifier to construct models. In our study, a cutoff threshold  used in the first phase is set to 50-th percentile and this cutoff threshold is used to identify higher metric values.


\noindent \paragraph{\textbf{UDP3.} } Zhang et al.~\cite{Zhang:ICSE2016} proposed a connectivity-based unsupervised method via spectral clustering.
They first used z-score to normalize
each metric, Then they used three steps for spectral clustering: (1) the first step was to calculate the Laplacian matrix
$L_{sym}$, (2) the second step was to perform the eigendecompositon on the matrix $L_{sym}$, (3) the third step was to divide
all the program modules into two clusters (i.e., the defective cluster and the non-defective cluster). To identify the
defective cluster, they used the following heuristic: for most of metrics, defective software modules generally had larger
values than non-defective software modules. Based on this heuristic, they used the average row sums of the normalized
metrics of each cluster. The cluster with larger average row sum was classified as the defective cluster and another cluster
was classified as the non-defective cluster.


\noindent \paragraph{\textbf{UDP4.} }
Zhou et al.~\cite{zhou2018far} proposed simple unsupervised methods (i.e., ManualDown and ManualUp) based on module size.  These methods are easy to implement and empirical results show these methods have a good performance.
In particular, ManualDown assumes a  module with larger LOC (lines of code) is more defect-prone,
while ManualUp assumes a  module with smaller LOC is more defect-prone.
In our study, the experimental setup for UDP4 is set as follows: based on the suggestions by Zhou et al.~\cite{zhou2018far}.
First for five groups of datasets (introduced in Section~\ref{chap4}) in our empirical studies, we choose the following metric as the metric LOC: LOC\_EXECUTABLE (The number of lines of executable code for a module) metric in the NASA dataset, executable\_loc (The number
of lines of executable code for am module) metric in the SOFTLAB dataset, ck\_oo\_numberOfLinesOfCode (Number of Lines of code) metric in the AEEEM
dataset, CountLineCode (Number of lines of code) metric in the ReLink dataset and loc (Number of lines of code) metric in the PROMISE dataset.
Second, in terms of non-effort-aware performance measures, we utilize the ManualDown method, since the top ranked modules by using this method may contain modules with high defect density. On the contrary, in terms of effort-aware performance measures, we utilize the ManualUp method.
Finally, we classified the top 50\% modules as defective and the remaining 50\% as
non-defective.

\noindent \paragraph{\textbf{UDP5.} }
However, when replicating  UDP4, we find when given the target project and the performance measure, LOC metric may not always achieve the best performance. For example, when considering the performance measure $\mathit{AUC}$, we identify the number of cases, which LOC metric can achieve the best performance and the number of cases, which other metrics (i.e., not LOC metric) can achieve the best performance. The results on different groups of datasets can be found in Table~\ref{RQ5analysis}.
 From Table~\ref{RQ5analysis}, we can find that LOC metric is not the best choice in most cases (30/34) in terms AUC performance measure. For example, When the target is Apache project, the metric CountLine can achieve the best performance.
  For other performance measures, we also draw the similar conclusion.
Therefore, we consider the fifth unsupervised methods, which is a variant of the method UDP4. In particular, given a specific performance measure, we choose the metric, which can achieve the best performance in the target project. We use UDP5 to denote this method.

\begin{table}[htbp]
  \centering
  \caption{Statistics of Optimal Metric  on Different Groups of Datasets in terms of AUC Performance Measure}
	\label{tab:1}
    \begin{tabular}{lccccc}
    \toprule
   Optimal Metric & \multicolumn{1}{l}{ReLink} & \multicolumn{1}{l}{AEEEM} & \multicolumn{1}{l}{SOFTLAB} & \multicolumn{1}{l}{PROMISE} & \multicolumn{1}{l}{NASA} \\
    \midrule
    LOC Metric  & 0     & 0     & 2     & 2     & 0 \\
    Other Metric & 3     & 5     & 3     & 8     & 11 \\
    \bottomrule
    \end{tabular}%
  \label{RQ5analysis}%
\end{table}%

\section{Experimental Setup}
\label{chap4}

\subsection{Research Questions}

In our empirical studies, we want to investigate the following four research questions.

\textbf{RQ1}: Do   these   HDP   methods   perform   significantly better than existing unsupervised methods in terms of non-effort-aware  performance  measures?

\textbf{RQ2}: Do   these   HDP   methods   perform   significantly better  than  existing  unsupervised  methods  in  terms  of effort-aware  performance  measures?

\textbf{RQ3}: How about the diversity analysis results on identifying  defective  modules  between  the  HDP  methods  and existing  unsupervised  methods?

\textbf{RQ4}:  Whether  the  HDP  methods  can  achieve  satisfactory  performance?

For the first three RQs, we  want to compare HDP methods with unsupervised methods from three different perspectives: non-effort-aware performance
measures (RQ1), effort-aware performance measures (RQ2)
and diversity analysis on identifying defective modules (RQ3).
In the last RQ (RQ4), we want to analyze the  ratio of these HDP methods with satisfactory performance
on different groups of datasets.

\subsection{Experimental Subjects}

In our empirical study, we choose publicly available datasets from five different groups (i.e., AEEEM, ReLink, PROMISE, NASA and SOFTLAB).
The diversity of these datasets from different groups can guarantee the generalization of our empirical results, since the software domain, utilized metrics, the granularity of extracted program modules are different in most of these groups. Moreover, these experimental subjects have been widely used
in previous HDP studies~\cite{nam2015heterogeneous,Jing2015Heterogeneous,li2018cost,li2017heterogeneous,LiASEJ2019} and the representative of these subjects can be guaranteed.

\noindent\textbf{AEEEM group.} The datasets from this group were gathered by D'Ambros et al.~\cite{Ambros:MSR2010}. These dataset includes 5 projects: EQ, JDT, LC,
ML and PDE. The granularity of extracted program modules is set to file. 61 metrics in total are used to measure the extracted modules.
In particular, 5 metrics are based on previous-defect information,  17 metrics are based on source code complexity, 5 metrics are based on entropy of change, 17 metrics are based on
churn of source code, and 17 metrics are based on entropy of source code.

\noindent\textbf{ReLink group.} The datasets from this group were gathered by Wu et al.~\cite{wu:FSE2011}
and the label information was  verified and
corrected in a manual way. The granularity of extracted program modules is set to file. 26 metrics are used to measure the extracted program
modules via the Understand tool. This dataset includes 3 data sets:  Apache, Safe and ZXing.

\noindent\textbf{PROMISE group.} The datasets from this group were gathered by Jureczko and Madeyski~\cite{jureczko2010towards} with the help of two tools: BugInfo and CKJM. The granularity of extracted program modules is set to class. 20 metrics based on code complexity are used to measure
the extracted program modules. More details of these metrics can be found in~\cite{jureczko2010towards}.

\noindent\textbf{NASA group.} The datasets from this group were original gathered by Menzies et al.~\cite{menzies2007data} and then were further cleaned by Shepperd
et al.~\cite{shepperd2013data}. The granularity of extracted program modules is set to function/method. Each project in NASA represents a NASA
software system or sub-system, which contains the corresponding defect-marking data and various static code metrics,
which include size, readability, complexity features and so on.

\noindent\textbf{SOFTLAB group.}
The datasets from this group were gathered by Turhan et al.~\cite{turhan:EMSE2009}. It consists of five projects, which are embedded
controller software for white goods.
The extracted modules are measured by 29 metrics. This dataset includes 6
projects (i.e., from AR1 to AR6).

The statistics of these datasets from different groups can be found in Table~\ref{data sets}. This table lists the name of the group, the name of the dataset, the
number of program modules and the number (percentage) of defective modules.
It is not hard to find datasets from different groups (i.e., gathered by different researchers) consider different metrics in most cases.
For example, NASA dataset has 21 to 37 metrics and AEEEM dataset has 61 metrics. The only common metric LOC (lines of codes) is used by both NASA dataset and AEEEM dataset.

\begin{table*}[htbp]
	\centering
	\caption{The Statistics of Datasets in Different Groups}
	\begin{tabular}{|c|c|c|c|c|c|}
		\hline
		\multirow{2}[0]{*} {Group} & \multirow{2}[0]{*}{Project Name} & \multicolumn{2}{c|}{Modules}   &   \multicolumn{1}{c|}{\multirow{2}[0]{*}{\# \newline{} Metrics}}  & \multicolumn{1}{c|}{\multirow{2}[0]{*}{Granularity}} \\	\cline{3-4}
		&       & \# Modules   & \# (\%) Defective Modules  &       &  \\
		\hline
		\multirow{5}[0]{*}{AEEEM} & EQ    & 324   & 129(39.81\%) & \multirow{5}[0]{*}{61} & \multirow{5}[0]{*}{Class} \\
		& JDT   & 997   & 206(20.66\%) &       &  \\
		& LC    & 691   & 64(9.26\%) &       &  \\
		& ML    & 1862  & 245(13.16\%) &       &  \\
		& PDE   & 1492  & 209(14.01\%) &       &  \\
		\hline
		\multirow{3}[0]{*}{ReLink} & Apache & 194   & 98(50.52\%) & \multirow{3}[0]{*}{26} & \multirow{3}[0]{*}{File} \\
		& Safe  & 56    & 22(39.29\%) &       &  \\
		& Zxing & 399   & 118(29.57\%) &       &  \\
		\hline
		\multirow{10}[0]{*}{PROMISE} & ant-1.3 & 125   & 20(16.00\%) & \multirow{10}[0]{*}{20} & \multirow{10}[0]{*}{Class} \\
		& arc   & 234   & 27(11.54\%) &       &  \\
		& camel-1.0 & 339   & 13(3.83\%) &       &  \\
		& poi-1.5 & 237   & 141(59.49\%) &       &  \\
		& redaktor & 176   & 27(15.34\%) &       &  \\
		& skarbonka & 45    & 9(20.00\%) &       &  \\
		& tomcat & 858   & 77(8.97\%) &       &  \\
		& velocity-1.4 & 196   & 147(75.00\%) &       &  \\
		& xalan-2.4 & 723   & 110(15.21\%) &       &  \\
		& xerces-1.2 & 440   & 71(16.14\%) &       &  \\
		\hline
		\multirow{11}[0]{*}{NASA} & cm1   & 344   & 42(12.21\%) & \multirow{5}[0]{*}{37} & \multirow{11}[0]{*}{Function} \\
		& mw1   & 264   & 27(10.23\%) &       &  \\
		& pc1   & 759   & 61(8.04\%) &       &  \\
		& pc3   & 1125  & 140(12.44\%) &       &  \\
		& pc4   & 1399  & 178(12.71\%) &       &  \\
		\cline{2-5}
		& jm1   & 9593  & 1759(18.34\%) & 21    &  \\\cline{2-5}
		& pc2   & 1585  & 16(1.01\%) & 36    &  \\\cline{2-5}
		& pc5   & 17001 & 503(2.96\%) & \multirow{2}[0]{*}{38} &  \\
		& mc1   & 9277  & 68(0.73\%) &       &  \\\cline{2-5}
		& mc2   & 127   & 44(34.65\%) & \multirow{2}[0]{*}{39} &  \\
		& kc3   & 200   & 36(18.00\%) &       &  \\
		\hline
		\multirow{5}[0]{*}{SOFTLAB} & ar1   & 121   & 9(7.44\%) & \multirow{5}[0]{*}{29} & \multirow{5}[0]{*}{Function} \\
		& ar3   & 63    & 8(12.70\%) &       &  \\
		& ar4   & 107   & 20(18.69\%) &       &  \\
		& ar5   & 36    & 8(22.22\%) &       &  \\
		& ar6   & 101   & 15(14.85\%) &       &  \\
		\hline
	\end{tabular}%
	\label{data sets}%
\end{table*}%

Since we only focus defect prediction across projects with heterogeneous metric set,
we do not conduct defect prediction across projects with the same metrics. For example, if we choose EQ in AEEEM datasets as the target project, there are 29 (=3+10+11+5) HDP combinations when considering the projects in other groups. Therefore, there are 876 (=$3\times 31+5\times 29+5\times 29+10\times 24+11\times 23$) HDP  combinations.
Notice some NASA datasets do no have the same metrics (there are five groups, each considers 37, 21, 36, 38 and 39 metrics respectively), therefore we also conduct HDP between  NASA datasets with different metrics. There are 86 (=$5\times 6+1\times 10+1\times 10+2\times 9+2\times 9$) HDP  combinations. In summary, we have  962 HDP combinations when considering these 34 datasets.

Notice we do not compare HDP with within-project defect prediction methods in our study, since this kind of comparison has been widely investigated in previous HDP studies to show   HDP is practical in practice~\cite{nam2015heterogeneous,nam2017heterogeneous,li2017heterogeneous}. Therefore, we do  not need to use 2-fold cross validation to evaluate the performance of different methods. On the contrary, in our study, we directly use the data in the source project as the training data and use the data in the target project as the test data.

\subsection{Performance Measures}

In our study, we mainly consider two categories of performance measures: non-effort-aware performance measures and effort-ware performance measures.

\subsubsection{Non-effort-aware Performance Measures}

For non-effort-aware performance measures, we mainly consider $\mathit{F1}$ and $\mathit{AUC}$.

In software defect prediction, if we treat defective modules as positive instances and non-defective modules as negative
instances, we can classify program modules into four types according to the actual type and predicted type of these
modules (i.e., true positive, false positive, true negative and false negative). We use $TP$, $FP$, $TN$ and $FN$ to denote the
number of true positives, false positives, true negatives, and false negatives respectively. The confusion matrix in the context of SDP
is shown in Table~\ref{confusionmatrix} and it forms the fundamental basis for computing threshold-dependent performance
measures. For this type of measures, if the predicted defective probability for a new module is larger than the given
threshold (0.5 is considered in most cases), this module will be classified as the defective module, otherwise, it will be classified
as the non-defective module. In this article, we introduce the following 3 threshold-dependent performance measures (i.e., $precision$,
$recall$ and $\mathit{F1}$) in sequence.

\begin{table}[htbp]
  \centering
  \caption{Confusion Matrix based on the Actual Type and Predicted Type of Program Modules}
    \begin{tabular}{ccc}
    \hline
    \multirow{2}[0]{*}{\textbf{Actual Program Type}} & \multicolumn{2}{c}{\textbf{Predicted Program Type}} \\
          & \textbf{Defective modules} & \textbf{Non-defective modules} \\\hline
     Defective modules & $TP$    & $FN$ \\
    Non-defective modules & $FP$    & $TN$ \\
   \hline
    \end{tabular}%
  \label{confusionmatrix}%
\end{table}%

\noindent\textbf{$\bm{precision}$ performance measure}. The performance measure $precision$ returns the ratio of the number of defective modules that are correctly classified
as defective to the number of modules that are classified as defective. This performance measure can be defined as:

\begin{equation}
precision = \frac{TP}{TP + FP}
\end{equation}

\noindent\textbf{$\bm{recall}$ performance measure}. The performance measure $recall$ returns the ratio of the number of defective modules that are correctly classified
as defective to the total number of defective modules. This performance measure can be defined as:

\begin{equation}
recall = \frac{TP}{TP + FN}
\end{equation}

\noindent\textbf{$\bm{F1}$ performance measure}. Since there exists trade-off between $precision$ and $recall$ in practice, in most cases, higher value of $precision$ means lower value of $recall$ and vice verse.  Therefore, it is hard to achieve both high $precision$ and $recall$ at the same time.
Here we use the performance measure $\mathit{F1}$, which is the harmonic mean between $precision$ and $recall$, to evaluate the overall performance of the trained models.
This performance measure can be defined as:

\begin{equation}
F1 = \frac{2\times precision \times recall}{precision + recall}
\end{equation}

\noindent\textbf{$\bm{AUC}$ performance measure}. However, previous studies~\cite{wang2018deep,tantithamthavorn:ICSE2016} have argued that these threshold-dependent performance
measures (such as $precision$, $recall$ and $\mathit{F1}$) are
problematic. For example, these performance measures depend on an manually selected
threshold and these measures are sensitive to class imbalanced problem, which exists in most of the gathered SDP
datasets.
Recently, researchers are more inclined  to use the AUC (Area Under
the receiver operator characteristic Curve) to measure the
discrimination power of the constructed models. $\mathit{AUC}$ is computed by measuring the area
under the curve, which plots the true positive rate against the
false positive rate, while varying the threshold  used
to determine whether a  module is classified as defective or
non-defective. The value of $\mathit{AUC}$ ranges between 0 (i.e., the worst performance)
and 1 (i.e., the best performance).  The higher the AUC value, the
better the prediction performance of the constructed models.
Notice a value of 0.5 indicates a performance, which is similar to the random method.

\subsubsection{Effort-aware Performance Measures.}
For effort-aware performance measures, we mainly consider $\mathit{ACC}$, $\mathit{P_{opt}}$, $\mathit{PMI@20\%}$ and $\mathit{IFA}$.

\noindent\textbf{$\bm{ACC}$ performance measure.} $\mathit{ACC}$ denotes the recall
of defective modules when expending 20\% of the entire efforts.

\noindent\textbf{$\bm{PMI@20\%}$ performance measure.} In the context of SDP,
$\mathit{PMI@20\%}$ returns proportion of modules inspected with only 20\% of the entire efforts. A higher $\mathit{PMI@20\%}$ value
indicates that, by only using 20\% of the entire efforts,
developers need to inspect more modules.
It means that the additional efforts required
due to context switches and additional communication
overhead among developer
~\cite{meyer:FSE2014}.

Therefore, when only using 20\% of the entire efforts, $\mathit{PMI@20\%}$ and $\mathit{ACC}$ evaluate different methods from two different perspectives.
We use a simple example to illustrate the difference between $\mathit{ACC}$ and $\mathit{PMI@20\%}$. Suppose there are 2000 modules in the project, of which 40 modules contain defects. If expending 20\% of the entire efforts based on the ranked list by a specific method, we can only inspect 600 modules, of which 10 modules are real defective modules. Then the $\mathit{ACC}$ of this method is 10/40 = 25\% and the $\mathit{PMI@20\%}$ of this method is 600/2000 = 30\%.

\noindent\textbf{$\bm{P_{opt}}$ performance measure.} $\mathit{P_{opt}}$
is the normalized version of the effort-aware performance indicator.
According to the previous study~\cite{kamei:ICSM2010}, $\mathit{P_{opt}}$ can be formally defined as:

\begin{equation}
  P_{opt}(m)=1-\frac{area(optimal)-area(m)}{area(optimal)-area(worst)}
\end{equation}

\begin{figure}[htbp]
\centering
\includegraphics[width=0.75\textwidth]{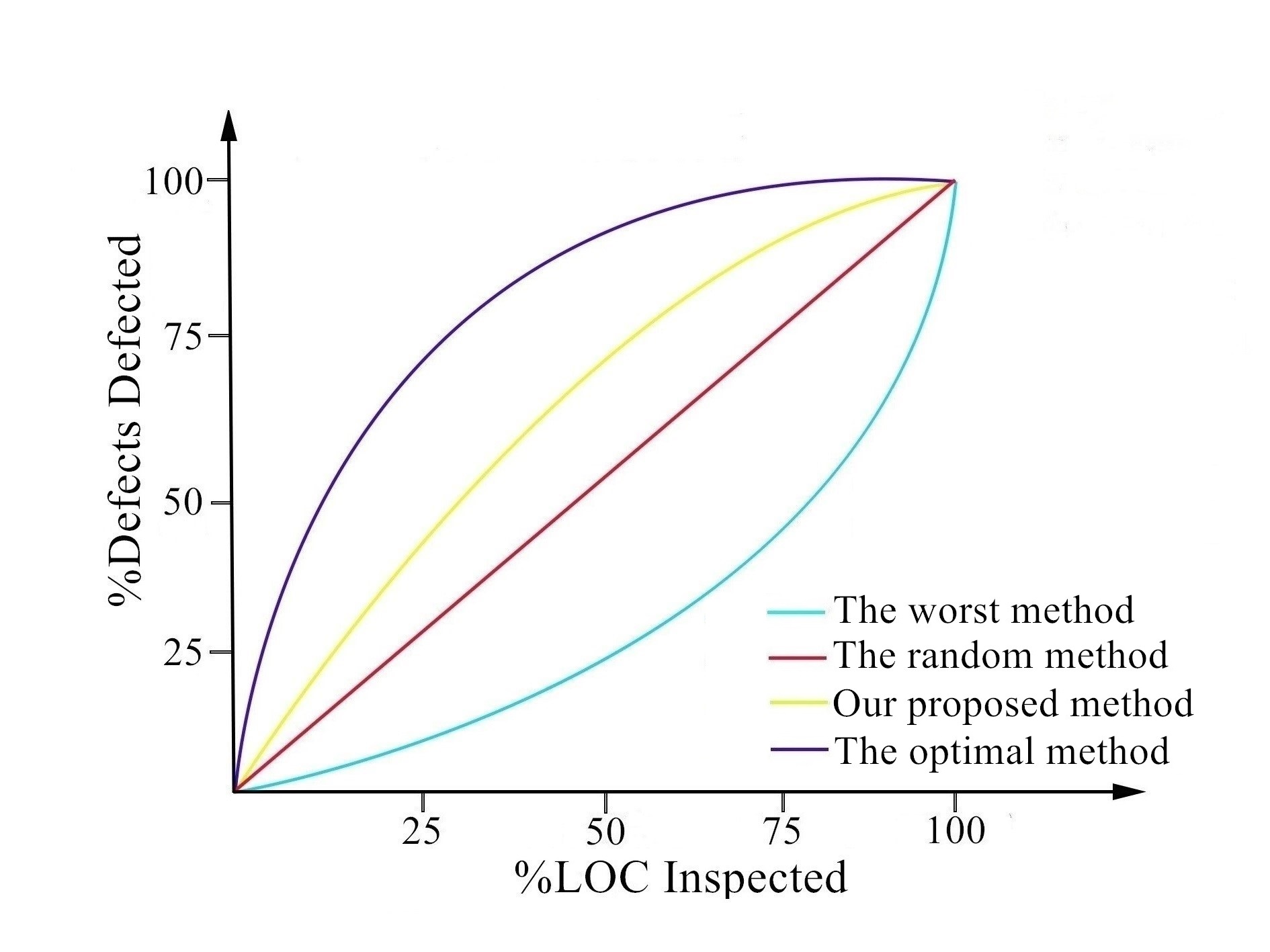}
\caption{The Illustrative Diagram for $\mathit{P_{opt}}$ Performance Measure}
 \label{POPTEXAMPLE}
\end{figure}

Here $area(m)$, $area(optimal)$ and $area(worst)$ are the area under the
curve corresponding to
a proposed prediction method, the optimal method, and the worst method, respectively.
For the optimal method, modules are sorted in the descendant order according to their actual defect density.
While for the worst method, modules are sorted in the ascendant order according to their actual defect density. The illustrative diagram can be found in Figure~\ref{POPTEXAMPLE}.

\noindent\textbf{$\bm{IFA}$ performance measure.}
$\mathit{IFA}$ returns the number of initial false alarms encountered before we
find the first defective module, which is inspired by research on automatic software fault localization~\cite{wong:TSE2016}.
A higher value of $\mathit{IFA}$ means more false positives (i.e., non-defective modules are predicted as defective modules) before detecting the first defective module and may have a non-ignorable impact on developers' confidence and tolerance
~\cite{parnin:ISSTA2011,kochhar:ISSTA2016}.

\section{Empirical Results}
\label{chap5}

\subsection{Results Analysis for RQ1}


\noindent\textbf{Motivation.}  In this RQ, we want to compare existing HDP methods with unsupervised methods in terms of non-effort-aware performance measures (i.e., $\mathit{F1}$ and $\mathit{AUC}$).

\noindent\textbf{Approach.}
To rank all the HDP methods and the unsupervised methods in terms of NPMs, we use Scott-Knott test. We use Scott-Knott test~\cite{jelihovschi2014scottknott} to rank all the HDP and UDP methods in terms of a specific  performance measure based on the results of all the empirical subjects. Scott-Knott test was recommended by Ghotra et al.~\cite{ghotra2015revisiting} when they compared different SDP methods. Since Scott-Knott test does not suffer
from the overlapping group issue in post hoc tests (such as Friedman-Nemenyi test). We can use Scott-Knott test to analyze whether some methods outperform others and it can generate a global ranking of all the methods. In particular, Scott-Knott test performs the grouping process in a recursive way. Firstly, Scott-Knott test uses a hierarchical cluster analysis method to partition all the methods into two ranks based on the mean performance in terms of a specific NPMs (such as $\mathit{AUC}$). Then, Scott-Knott test is recursively executed again in each rank to further divide the ranks if the divided ranks are significantly different. The test will terminate when the ranking can   no longer be divided into statistically different rankings.

We use Wilcoxon signed-rank test to analyze whether the performance difference
between two methods are statistically significant. We also use the
Benjamini-Hochberg (BH) procedure to adjust $p$ values if we make multiple
comparisons. Then if the test result shows a significant difference, we compute Cliff's
$\delta$~\cite{benjamini:1995}, which is a non-parametric effect size measure, to examine whether the magnitude
of the difference between two  models is substantial or not. In summary,
a  method performs significantly better or worse than another method, if BH corrected $p$ value is less than 0.05 and the effectiveness level is not negligible
based on Cliff's $\delta$ (i.e., $|\delta| \geq 0.147$). While the difference between two methods is not significant,
if $p$ value is not less than 0.05 or $p$-value is less than 0.05 and the effectiveness level is negligible (i.e., $|\delta| < 0.147$).

\noindent\textbf{Results.}
Since there exist some cross-project defect prediction combinations HDP1 cannot success (i.e., it cannot find  a group of matched
metrics between the source project and the target project according to the description of HDP1).
Therefore, we perform  Scott-Knott test in two scenarios. In the first scenario (i.e., Scenario1), we do not consider HDP1 and this scenario will consider all the HDP prediction combinations.
In the second scenario (i.e., Scenario2), we consider HDP1 and this scenario will only consider all the HDP combinations HDP1 can success.
In our study, the number of all the HDP combinations is 962, and the method HP1 can success in 623 HDP combinations.

\begin{figure}[htbp]
\centering
\subfigure[Scenario1]{
\begin{minipage}[t]{\linewidth}
\centering
\includegraphics[width=1\linewidth]{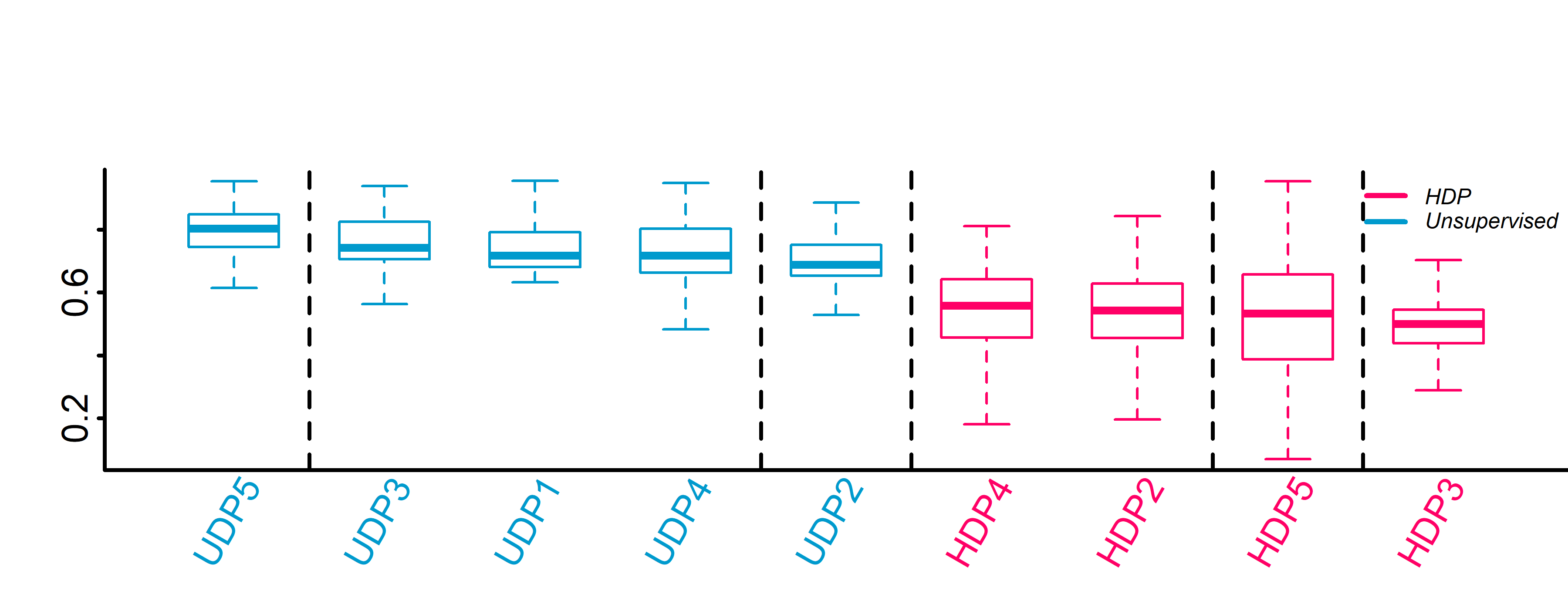}
\end{minipage}%
}%
\quad
\subfigure[Scenario2]{
\begin{minipage}[t]{\linewidth}
\centering
\includegraphics[width=1\linewidth]{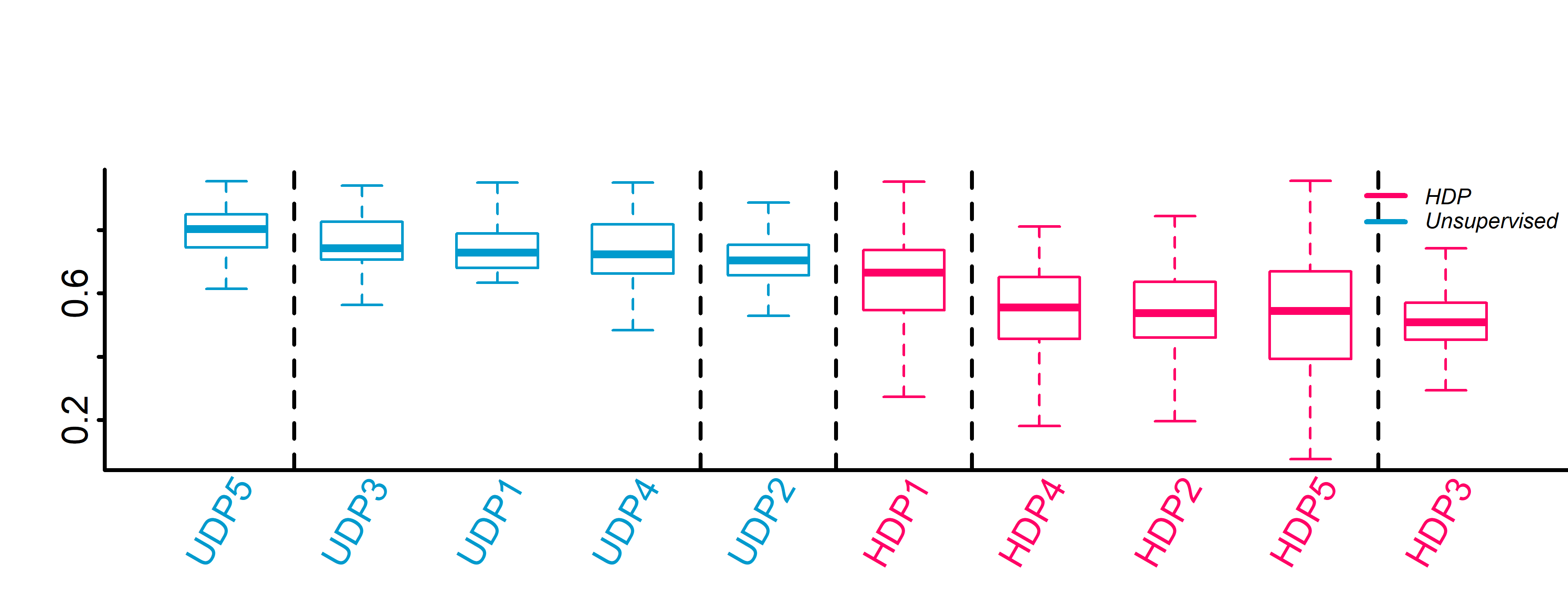}
\end{minipage}%
}%
\centering
\caption{Scott-Knott Test Results in terms of AUC Performance Measure}
 \label{RQ1AUC}
\end{figure}

The comparison results in terms of $\mathit{AUC}$ based on the Scott-Knott test in two scenarios on different groups of datasets can be found in Figure~\ref{RQ1AUC}.
In two subfigures,
the
dotted lines represent groups divided by using the Scott-Knott test.
All methods are ordered based on their mean ranks.  The blue label denotes unsupervised
methods and the red label denotes supervised methods. From these two subfigures, we can find all the 5 unsupervised methods can significantly perform  better   than supervised methods in terms of $\mathit{AUC}$ performance measure.
Among these 5 unsupervised methods, UM5 can achieve the best performance.
We also report win/tie/loss result of comparing the HDP methods and the unsupervised methods in terms of $\mathit{AUC}$ when given the target project. Supposing the method1 $vs.$ method2, ``Win'' means the number of HDP combinations the method1 can perform significantly better than the method2. ``Tie'' means the number of HDP combinations the performance between the method1 and the method2 has no statistical significance. ``Loss''  means the number of HDP combinations the method1 can perform significantly worse than the method2. The results can be found in Table~\ref{RQ1WTLAUC}. From this table, we find: (1) In the scenario1, the UDP methods can win HDP2, HDP3, HDP4 and HDP5 at least 31, 31, 27 and 29 times. (2) In the scenario2, the UDP methods can win   HDP2, HDP3, HDP4 and HDP5 at least 28, 31, 25 and 27. When compared to HDP1, UDP methods except for UDP2 can win at least 21 times.
 These results show the UDP methods can perform significantly better than the HDP methods in majority of cases when considering $\mathit{AUC}$.

\begin{table}
	\caption{Win/Tie/Loss Analysis in terms of $\mathit{AUC}$ Performance Measure}
	\centering
	\subtable[Scenario1]{
		\begin{tabular}{cccccc}\hline
		Method Name & UDP1 & UDP2 & UDP3 & UDP4 &UDP5\\
		\hline
		  HDP2 & 2/1/31 & 2/1/31 & 2/1/31 & 3/0/31 & 0/0/34 \\
    HDP3 & 3/0/31 & 2/1/31 & 2/1/31 & 3/0/31 & 0/0/34 \\
    HDP4 & 3/0/31 & 3/4/27 & 2/2/30 & 3/2/29 & 0/0/34 \\
    HDP5 & 0/2/32 & 1/4/29 & 1/1/32 & 0/4/30 & 0/0/34 \\

		\hline
	\end{tabular}
		}
	\\
	\subtable[Scenario2]{
		\begin{tabular}{cccccc}\hline
		Method Name & UDP1 & UDP2 & UDP3 & UDP4 &UDP5\\
		\hline
		    HDP1 & 3/10/21 & 4/19/11 & 1/11/22 & 5/7/22 & 0/0/34 \\
    HDP2 & 2/1/31 & 2/4/28 & 2/2/30 & 3/2/29 & 0/0/34 \\
    HDP3 & 3/0/31 & 2/1/31 & 2/1/31 & 3/0/31 & 0/0/34 \\
    HDP4 & 3/1/30 & 3/6/25 & 2/3/29 & 3/3/28 & 0/0/34 \\
    HDP5 & 0/2/32 & 1/6/27 & 1/2/31 & 0/6/28 & 0/0/34 \\
		\hline
	\end{tabular}
	
	}
		\label{RQ1WTLAUC}
\end{table}

\begin{figure}[htbp]
\centering
\subfigure[Scenario1]{
\begin{minipage}[t]{\linewidth}
\centering
\includegraphics[width=1\linewidth]{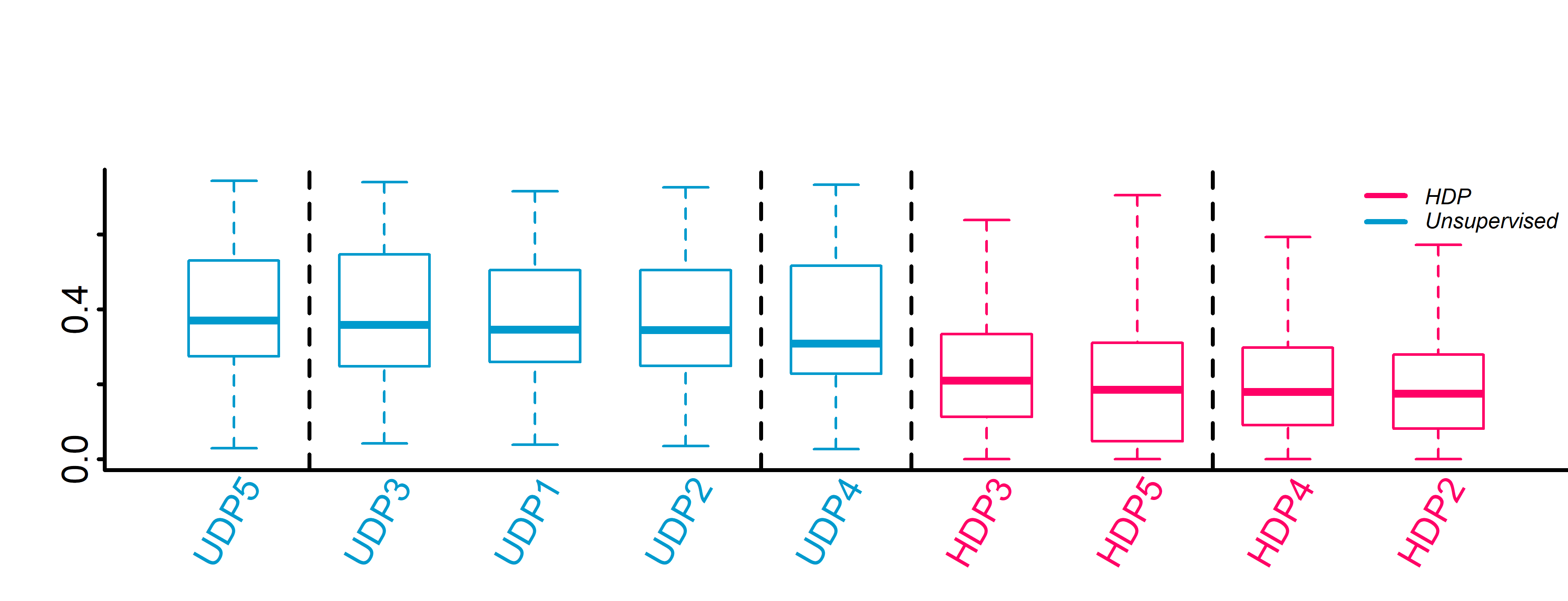}
\end{minipage}%
}%
\quad
\subfigure[Scenario2]{
\begin{minipage}[t]{\linewidth}
\centering
\includegraphics[width=1\linewidth]{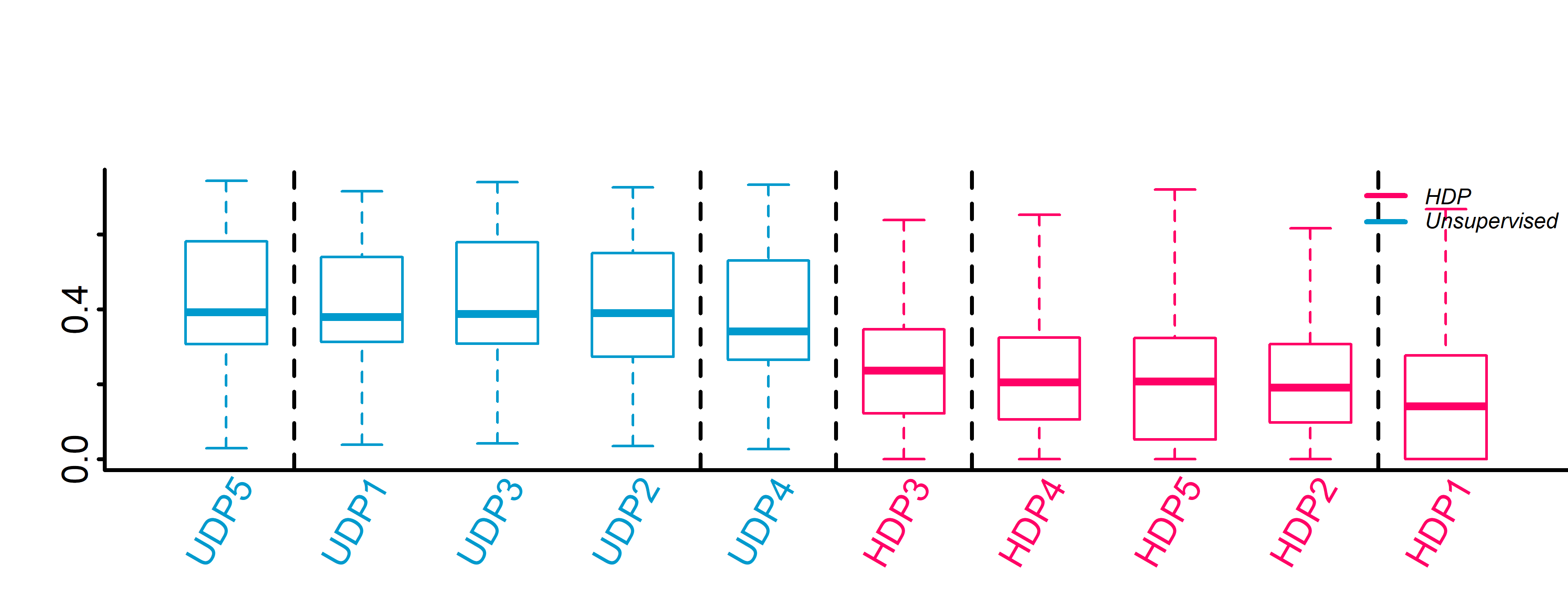}
\end{minipage}%
}%
\centering
\caption{Scott-Knott Test Results in terms of $\mathit{F1}$ Performance Measure}
 \label{RQ1F1}
\end{figure}

The comparison results in terms of $\mathit{F1}$ based on the Scott-Knott test in two scenarios on different groups of datasets can be found in Figure~\ref{RQ1F1}.  From these two subfigures, we can find all the 5 unsupervised methods can  significantly perform better  than the HDP methods in terms of $\mathit{F1}$ performance measure.
Among these 5 unsupervised methods, UM5 can achieve the best performance.
Win/Tie/Loss result of comparing the HDP methods and the unsupervised methods in terms of $\mathit{F1}$ can be found in Table~\ref{RQ1WTLF1}.
From this table, we find: (1) In the scenario1, the UDP methods can win HDP2, HDP3, HDP4 and HDP5 at least 31, 29, 31 and 27 times. (2) In the scenario2, the UDP methods can win HDP1,  HDP2, HDP3, HDP4 and HDP5 at least 26, 30, 29, 30, 30 and 26 times. These results show the UDP methods can perform significantly better than the HDP methods in majority of cases when considering $\mathit{F1}$.

\begin{table}
	\caption{Win/Tie/Loss Analysis in terms of  $\mathit{F1}$ Performance Measure}
	\centering
	\subtable[Scenario1]{
		\begin{tabular}{cccccc}\hline
		Method Name & UDP1 & UDP2 & UDP3 & UDP4 &UDP5\\
		\hline
		  HDP2 & 0/2/32 & 0/1/33 & 1/2/31 & 0/3/31 & 0/0/34 \\
    HDP3 & 2/1/31 & 3/0/31 & 3/2/29 & 3/0/31 & 0/0/34 \\
    HDP4 & 0/2/32 & 1/2/31 & 1/2/31 & 1/2/31 & 0/0/34 \\
    HDP5 & 0/1/33 & 0/4/30 & 1/6/27 & 2/4/28 & 1/1/32 \\

		\hline
	\end{tabular}
		}
	\\
	\subtable[Scenario2]{
		\begin{tabular}{cccccc}\hline
		Method Name & UDP1 & UDP2 & UDP3 & UDP4 &UDP5\\
		\hline
    HDP1 & 0/6/28 & 1/5/28 & 0/6/28 & 1/7/26 & 1/4/29 \\
    HDP2 & 0/2/32 & 0/3/31 & 1/3/30 & 0/3/31 & 0/0/34 \\
    HDP3 & 2/1/31 & 2/2/30 & 3/2/29 & 3/0/31 & 0/0/34 \\
    HDP4 & 0/1/33 & 0/3/31 & 1/3/30 & 0/3/31 & 0/0/34 \\
    HDP5 & 0/4/30 & 0/6/28 & 1/5/28 & 0/8/26 & 0/3/31 \\
		\hline
	\end{tabular}
	}
		\label{RQ1WTLF1}
\end{table}

\subsection{Results Analysis for RQ2}

\noindent\textbf{Motivation.}  In this RQ, we want to compare existing HDP methods with the unsupervised methods in terms of effort-aware performance measures.

\noindent\textbf{Approach.}
To answer this RQ, we also want to use  Scott-Knott test~\cite{jelihovschi2014scottknott} to rank all the HDP and UDP methods in terms of a specific EPM.

\begin{figure}[htbp]
\centering
\subfigure[Scenario1]{
\begin{minipage}[t]{\linewidth}
\centering
\includegraphics[width=1\linewidth]{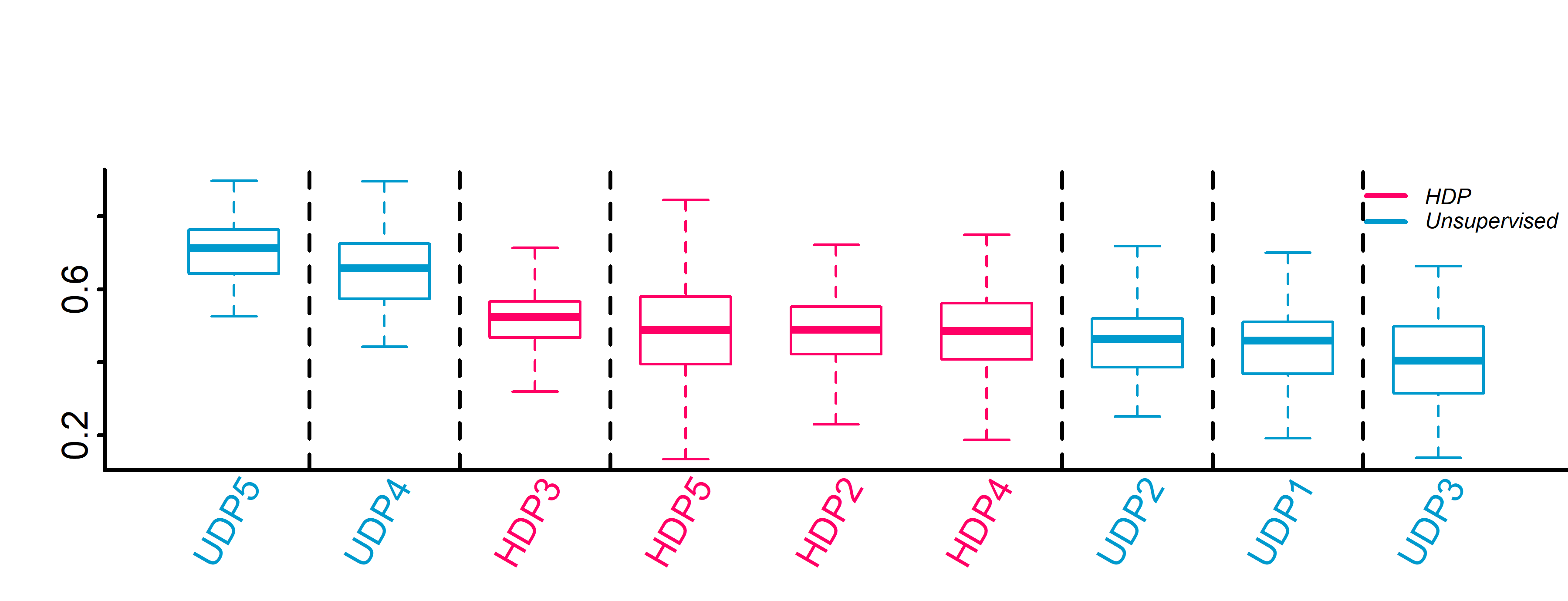}
\end{minipage}%
}%
\quad
\subfigure[Scenario2]{
\begin{minipage}[t]{\linewidth}
\centering
\includegraphics[width=1\linewidth]{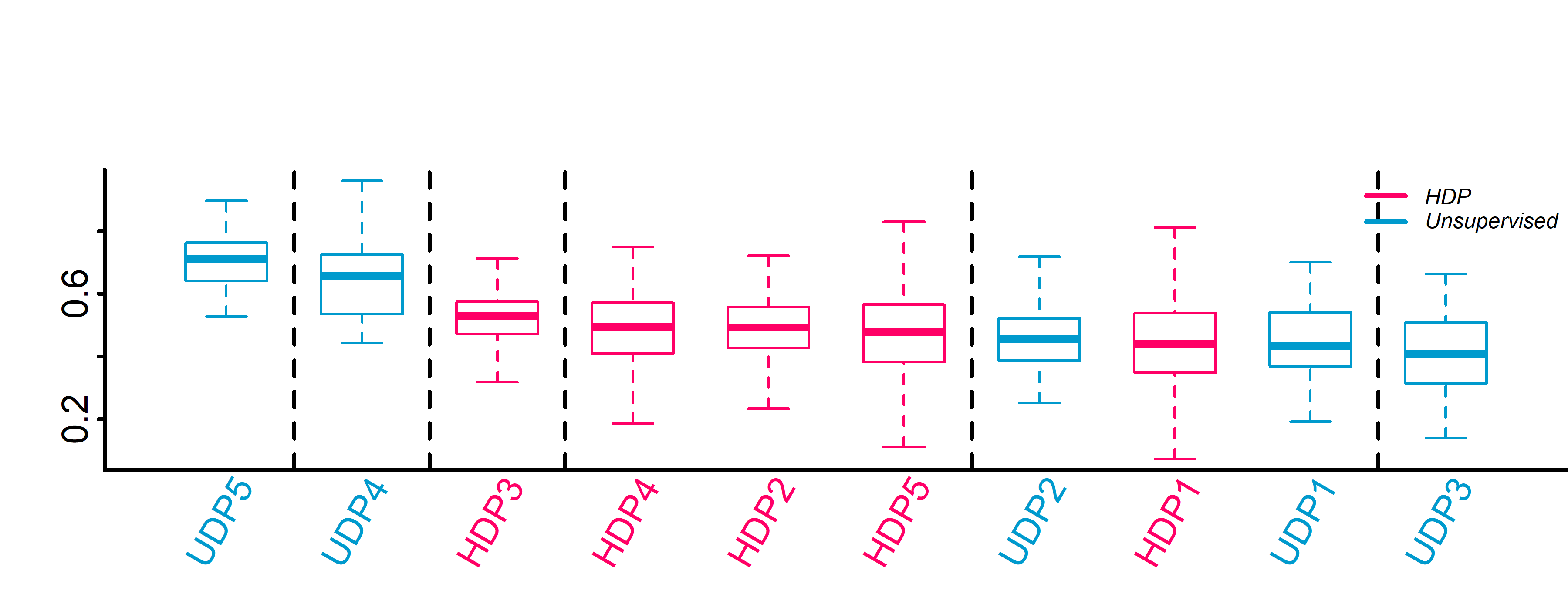}
\end{minipage}%
}%
\centering
\caption{Scott-Knott Test Results in terms of $\mathit{P_{opt}}$ Performance Measure}
 \label{RQ2POPT}
\end{figure}

\noindent\textbf{Results.}
The Scott-Knott test results in terms of $\mathit{P_{opt}}$ can be found in  Figure~\ref{RQ2POPT}.
From these two subfigures, we can find the   unsupervised methods UM5 and UM4  can perform significantly better than the supervised methods in terms of $\mathit{P_{opt}}$ performance measure. Win/Tie/Loss result of comparing the HDP methods and the unsupervised methods in terms of $\mathit{P_{opt}}$ can be found in Table~\ref{RQ2WTLpopt}.
From this table, we find: (1) In the scenario1, the UDP methods UDP4 and UDP5 can win HDP methods at least 27 and 31 times. However, the UDP methods UDP1, UDP2 and UDP3 can only win HDP methods at most 11, 11 and 7 times. (2) In the scenario2, the UDP methods UDP4 and UDP5 can win HDP methods at least 25  and 31 times. However, the UDP methods UDP1, UDP2 and UDP3 can only win HDP methods at most 10, 16 and 6 times.
These results show the UDP methods UDP4 and UDP5 can perform significantly better than the HDP methods in majority of cases when considering $\mathit{P_{opt}}$.


\begin{table}
	\caption{Win/Tie/Loss Analysis in terms of $\mathit{P_{opt}}$ Performance Measure}
	\centering
	\subtable[Scenario1]{
		\begin{tabular}{cccccc}\hline
		Method Name & UDP1 & UDP2 & UDP3 & UDP4 &UDP5\\
		\hline
		 HDP2 & 24/1/9 & 18/5/11 & 25/4/5 & 2/2/30 & 0/1/33 \\
    HDP3 & 21/2/11 & 22/3/9 & 27/1/6 & 6/0/28 & 1/2/31 \\
    HDP4 & 22/4/8 & 21/3/10 & 25/3/6 & 2/3/29 & 0/1/33 \\
    HDP5 & 16/9/9 & 12/14/8 & 20/7/7 & 2/5/27 & 0/0/34 \\

		\hline
	\end{tabular}
		
	}
	\\
	\subtable[Scenario2]{
		\begin{tabular}{cccccc}\hline
		Method Name & UDP1 & UDP2 & UDP3 & UDP4 &UDP5\\
		\hline
    HDP1 & 8/20/6 & 9/9/16 & 15/14/5 & 4/4/26 & 1/1/32 \\
    HDP2 & 24/4/6 & 17/8/9 & 28/1/5 & 2/4/28 & 0/1/33 \\
    HDP3 & 21/3/10 & 22/3/9 & 26/2/6 & 6/0/28 & 1/2/31 \\
    HDP4 & 21/5/8 & 19/7/8 & 23/5/6 & 2/4/28 & 0/1/33 \\
    HDP5 & 14/12/8 & 6/21/7 & 16/12/6 & 1/8/25 & 0/0/34 \\
		\hline
	\end{tabular}
	}
	\label{RQ2WTLpopt}
\end{table}

\begin{figure}[htbp]
\centering
\subfigure[Scenario1]{
\begin{minipage}[t]{\linewidth}
\centering
\includegraphics[width=1\linewidth]{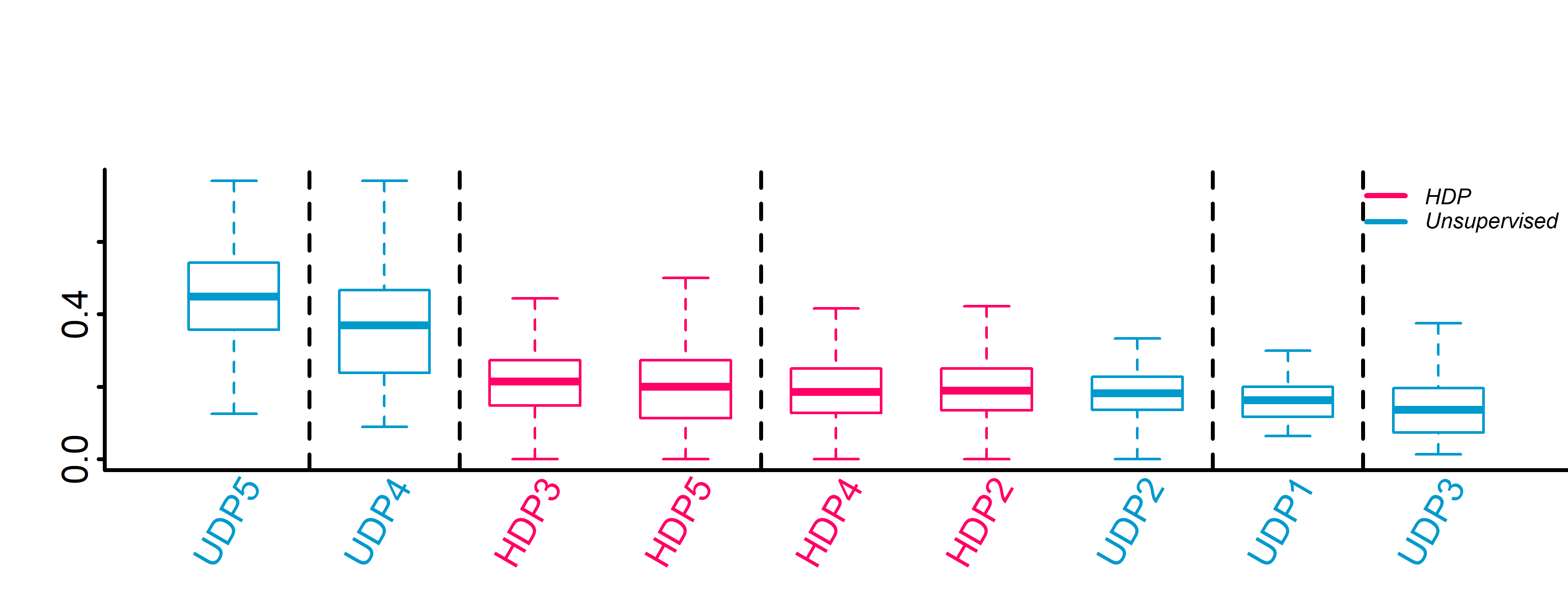}
\end{minipage}%
}%
\quad
\subfigure[Scenario2]{
\begin{minipage}[t]{\linewidth}
\centering
\includegraphics[width=1\linewidth]{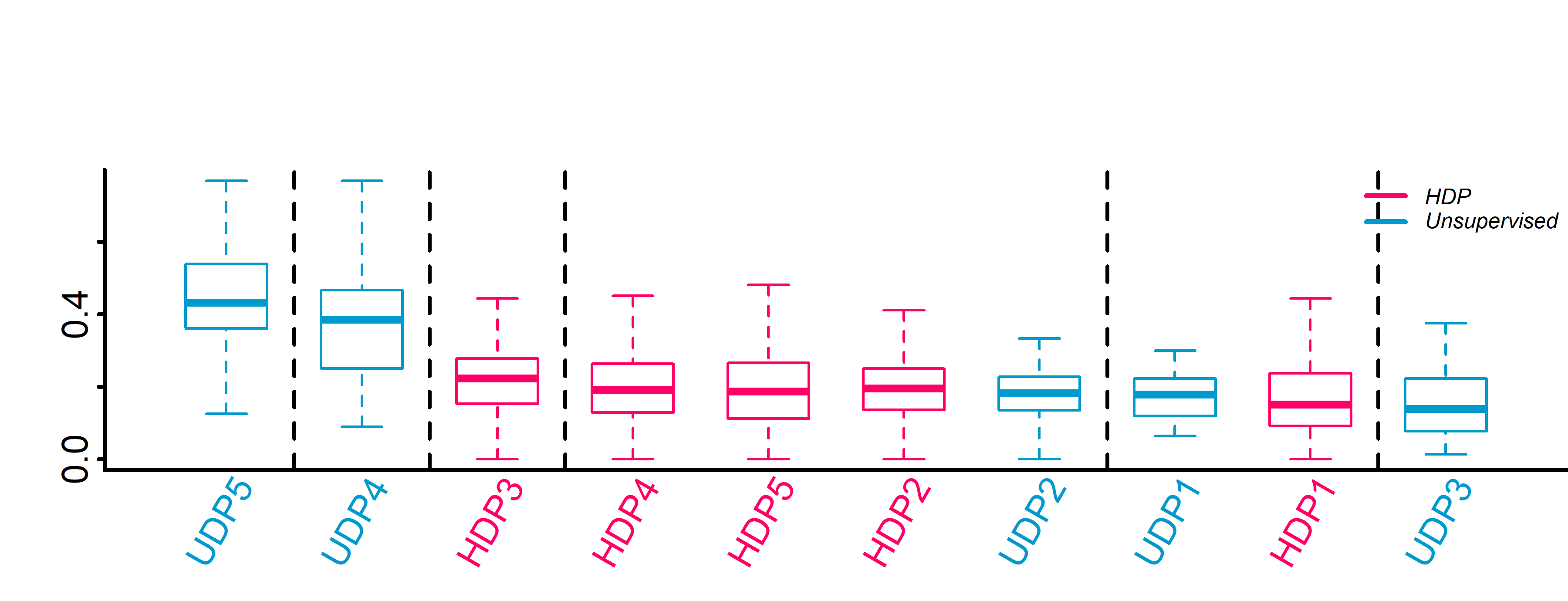}
\end{minipage}%
}%
\centering
\caption{Scott-Knott Test Results in terms of $\mathit{ACC}$ Performance Measure}
 \label{RQ2ACC}
\end{figure}

The Scott-Knott test results in terms of $\mathit{ACC}$ can be found in  Figure~\ref{RQ2ACC}. From these two subfigures, we can find  similar to the performance measure $\mathit{P_{opt}}$, the unsupervised methods UM5 and UM4 can also significantly perform better than the supervised methods in terms of $\mathit{ACC}$ performance measure.
Win/Tie/Loss result of comparing the HDP methods and the unsupervised methods in terms of $\mathit{ACC}$ can be found in Table~\ref{RQ2WTLACC}.
From this table, we find: (1) In the scenario1, the UDP methods UDP4 and UDP5 can win HDP methods at least 25 and 32 times. However, the UDP methods UDP1, UDP2 and UDP3 can only win HDP methods at most 12, 14 and 7 times. (2) In the scenario2, the UDP methods UDP4 and UDP5 can win HDP methods at least 25  and 32 times. However, the UDP methods UDP1, UDP2 and UDP3 can only win HDP methods at most 12, 18 and 8 times.
These results show the UDP methods UDP4 and UDP5 can perform significantly better than the HDP methods in majority of cases when considering $\mathit{ACC}$.


\begin{table}
	\caption{Win/Tie/Loss Analysis in terms of $\mathit{ACC}$ Performance Measure}
	\centering
	\subtable[Scenario1]{
		\begin{tabular}{cccccc}\hline
		Method Name & UDP1 & UDP2 & UDP3 & UDP4 &UDP5\\
		\hline
		 HDP2  & 18/7/9 & 10/10/14 & 23/5/6 & 3/6/25 & 1/0/33 \\
    HDP3  & 24/4/6 & 19/4/11 & 25/4/5 & 6/3/25 & 2/0/32 \\
    HDP4  & 18/4/12 & 12/10/12 & 23/4/7 & 4/2/28 & 1/0/33 \\
    HDP5  & 14/13/7 & 7/20/7 & 18/9/7 & 3/6/25 & 0/2/32 \\

		\hline
	\end{tabular}
	}
	\\
	\subtable[Scenario2]{
		\begin{tabular}{cccccc}\hline
		Method Name & UDP1 & UDP2 & UDP3 & UDP4 &UDP5\\
		\hline
    HDP1  & 5/17/12 & 6/10/18 & 12/14/8 & 3/6/25 & 1/0/33 \\
    HDP2  & 17/11/6 & 10/14/10 & 22/7/5 & 2/7/25 & 1/0/33 \\
    HDP3  & 25/3/6 & 19/6/9 & 26/3/5 & 6/3/25 & 1/1/32 \\
    HDP4  & 16/9/9 & 10/14/10 & 22/6/6 & 3/4/27 & 1/0/33 \\
    HDP5  & 13/13/8 & 5/20/9 & 16/11/7 & 2/7/25 & 0/2/32 \\
		\hline
	\end{tabular}
	}
	\label{RQ2WTLACC}
\end{table}

\begin{figure}[htbp]
\centering
\subfigure[Scenario1]{
\begin{minipage}[t]{\linewidth}
\centering
\includegraphics[width=1\linewidth]{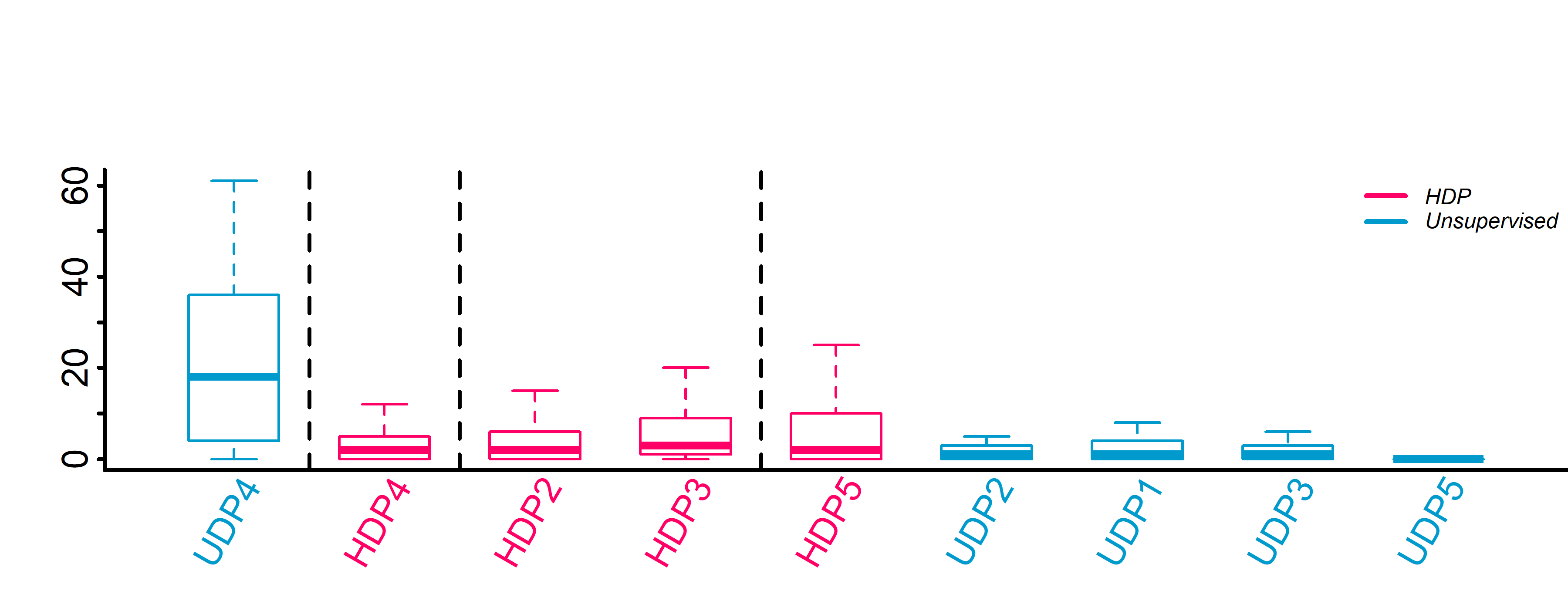}
\end{minipage}%
}%
\quad
\subfigure[Scenario2]{
\begin{minipage}[t]{\linewidth}
\centering
\includegraphics[width=1\linewidth]{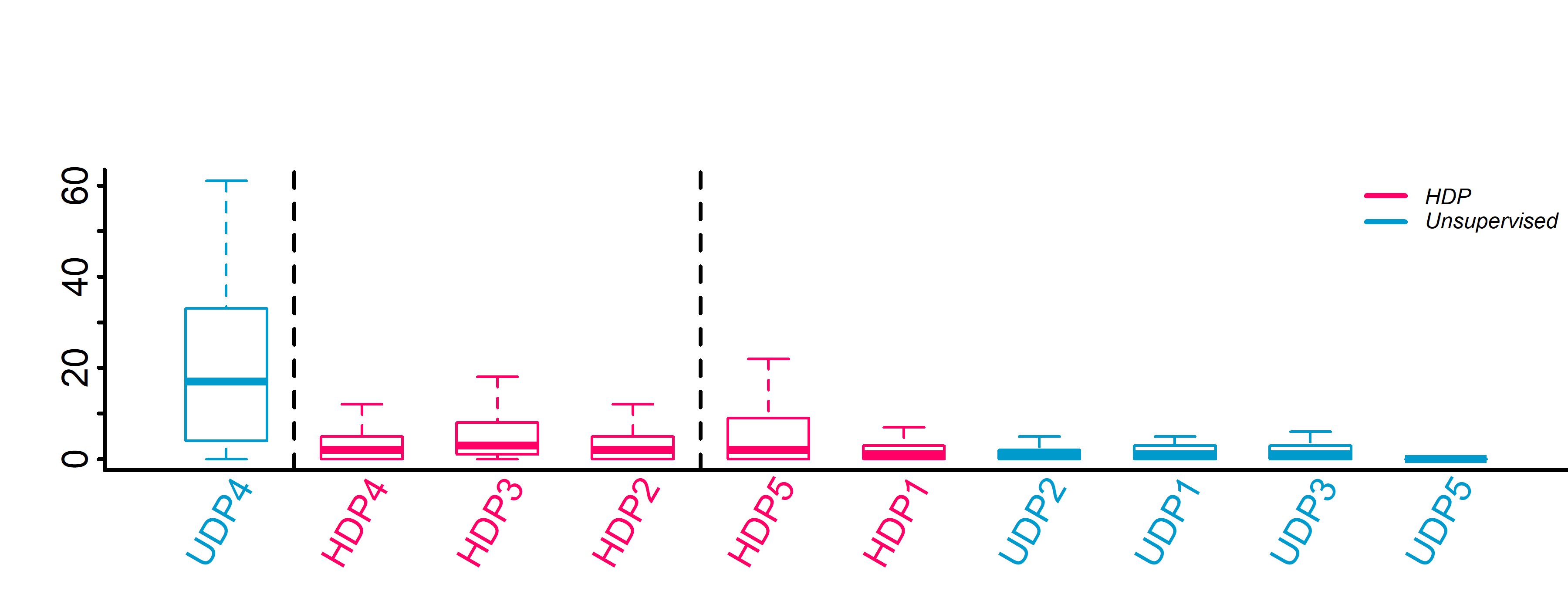}
\end{minipage}%
}%
\centering
\caption{Scott-Knott Test Results in terms of $\mathit{IFA}$ Performance Measure}
 \label{RQ2IFA}
\end{figure}

The Scott-Knott test results in terms of $\mathit{IFA}$ can be found in  Figure~\ref{RQ2IFA}. From these two figures, we can find the unsupervised method UM4 has highest $\mathit{IFA}$ value, which is in consistent with the findings by Huang et al.~\cite{huang2018revisiting}. However, we surprise find that the supervised methods have higher $\mathit{IFA}$ value than the remaining unsupervised methods.


\begin{figure}[htbp]
\centering
\subfigure[Scenario1]{
\begin{minipage}[t]{\linewidth}
\centering
\includegraphics[width=1\linewidth]{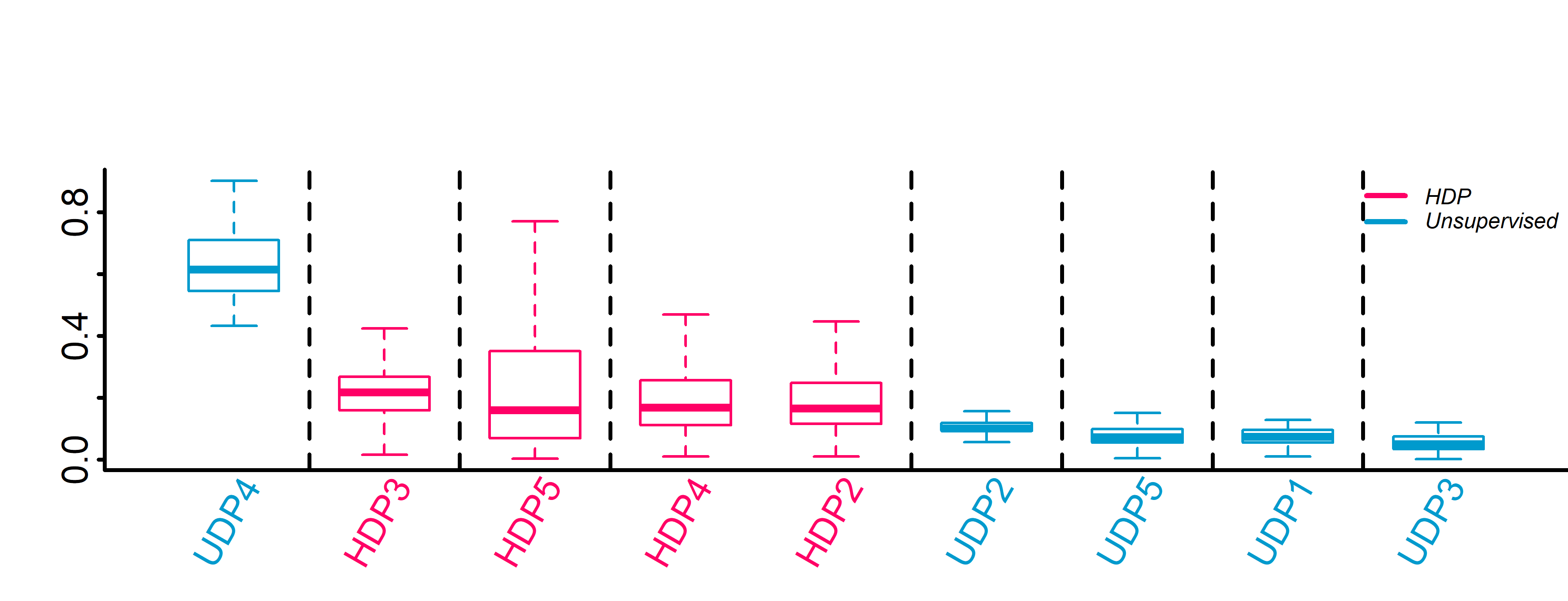}
\end{minipage}%
}%
\quad
\subfigure[Scenario2]{
\begin{minipage}[t]{\linewidth}
\centering
\includegraphics[width=1\linewidth]{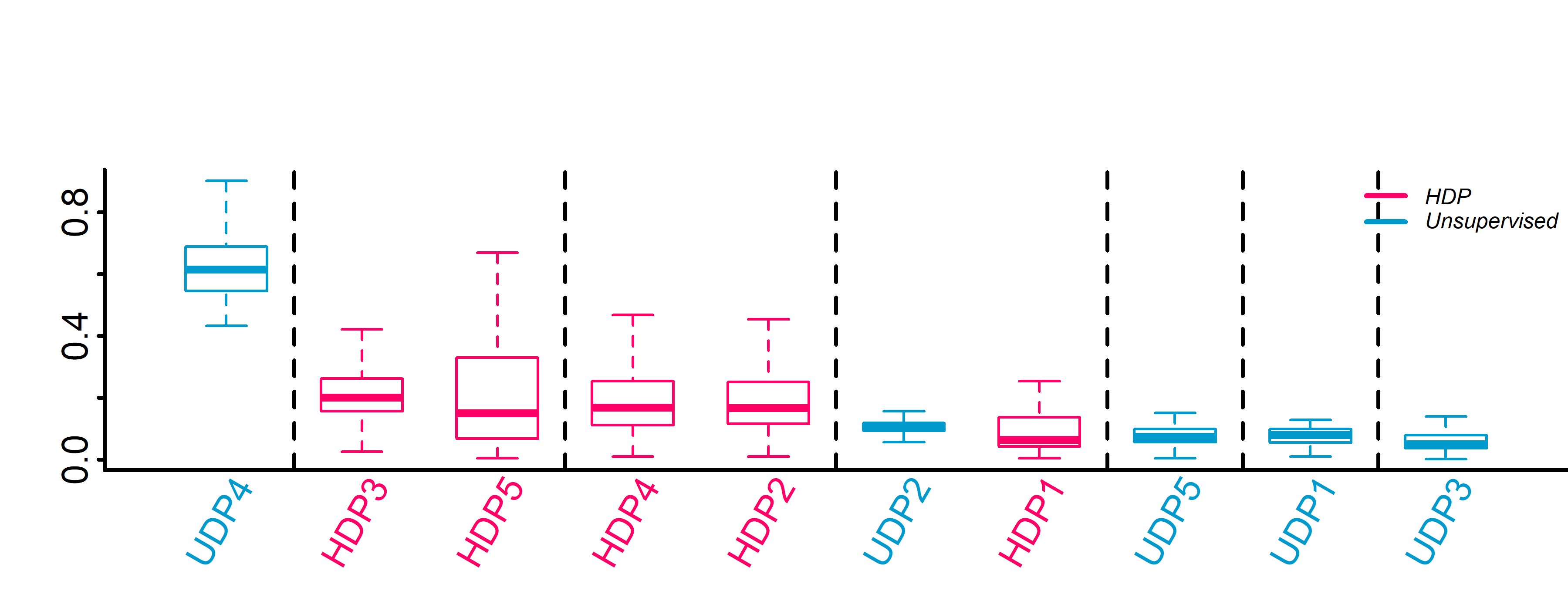}
\end{minipage}%
}%
\centering
\caption{Scott-Knott Test Results in terms of $PMI@20\%$ Performance Measure}
 \label{RQ2PCI}
\end{figure}

The Scott-Knott test results in terms of $PMI@20\%$ can be found in  Figure~\ref{RQ2PCI}. From these two subfigures, similar to $\mathit{IFA}$, we can find the unsupervised method UM4 has highest $PMI@20\%$ value,  which is also in consistent with the findings by Huang et al.~\cite{huang2018revisiting}.
However, we also surprise find that supervised methods have higher $PMI@20\%$ value than the remaining unsupervised methods.

\subsection{Results Analysis for RQ3}

\noindent\textbf{Motivation.}
In previous two RQs, we compare and rank the performance of the HDP methods and the unsupervised methods in terms of effort-aware performance measures and non-effort-aware performance measures.
In this RQ, we want to perform a deep analysis  on identifying defective modules between different types of methods.

\noindent\textbf{Approach.}
To evaluate whether different  methods (i.e., the HDP methods or the unsupervised methods) result in distinct predictions on defective modules, we want to test the following null hypothesis:

\textbf{Ho: Both the methods $M1$ and $M2$ can identify similar defective modules in the target project.}

We use McNemar's test~\cite{dietterich1998approximate}, which is used in the previous study~\cite{bowes2018software},
to perform diversity analysis on identifying defective modules
with the 95\% confidence level  between different  methods. McNemar's test is a non-parametric test and it does not need any assumption on the distribution of a subject variable.

To perform McNemar's test, we need to construct a contingency matrix based on the prediction results by two  different methods (i.e., $M1$ and $M2$) and this matrix is shown in Table~\ref{contingency matrix}. In this contingency matrix, $N_{cc}$ denotes the number of
 modules that both  methods can achieve correct predictions. $N_{cw}$ denotes the number of defective modules
that the method $M1$  can achieve a correct prediction while the method $M2$ cannot achieve a correct prediction. $N_{wc}$ denotes the number of defective modules that the  method $M2$  can achieve a correct prediction while the  method $M1$ cannot achieve a correct prediction. Finally $N_{ww}$ denotes the number of defective modules that both  methods cannot achieve correct predictions.

\begin{table}[htb]
  \centering
  \caption{A Contingency Matrix based on the Prediction Results on the Target Project by two  Methods $M1$ and $M2$ }
    \begin{tabular}{l c c}
    \toprule
     $M1$ $vs.$ $M2$ & Correct Predictions & Wrong Predictions \\
    \midrule
    Correct Predictions & $N_{cc}$   &  $N_{cw}$  \\
    Wrong Predictions &   $N_{wc}$  &   $N_{ww}$  \\
    \bottomrule
    \end{tabular}%
  \label{contingency matrix}%
\end{table}%

In our empiric studies, we use mcnemar function in statsmodel provided by R package exact$2\times2^6$ to perform McNemar's test. If $p$  value is smaller than 0.05, we will reject the null hypothesis Ho (i.e., The defective modules in the target project identified by these two  methods $M1$ and $M2$ are almost the same). We use an artificially constructed illustrative example in Table \ref{diversity analysis} to show the rationality of the diversity analysis method. In this table, there are 10 defective modules (i.e., {m0, m1, $\cdots$, m9}). The prediction results of three different methods (i.e., Method1, Method2 and Method3) can be found in the last three columns. Here 1 means  this program module is predicted as the defective module and 0 means  this program module is predicted as the non-defective module by the corresponding  method. Based on the McNemar's test, the $p$ value of Method1 $vs.$ Method2 is 0.0196 ($<$0.05) and this means these two methods can almost identify distinct defective modules. While the $p$ value of Method1 $vs.$ Method3 is 0.4142 ($>$0.05) and this means these two methods almost identify the same defective modules.

\begin{table}[htb]
  \centering
  \caption{An Illustrative Example to Show the Rationality of the Diversity Analysis Method used in Our Study}
    \begin{tabular}{c c c c}
    \toprule
    Defective Module & Method1 & Method2 & Method3 \\
    \midrule
    m0 & 0  &  1  &  0  \\
    m1 & 0  &  1  &  0  \\
    m2 & 0  &  1  &  1 \\
    m3 & 0  &  1  &  1 \\
    m4 & 0  &  1  &  0 \\
    m5 & 1  &  0  &  0 \\
    m6 & 1  &  1  &  0 \\
    m7 & 0  &  1  &  1 \\
    m8 & 0  &  1  &  1 \\
    m9 & 0  &  1  &  0 \\
    \bottomrule
    \end{tabular}%
  \label{diversity analysis}%
\end{table}%

Furthermore, we also analyze the number of defective modules in the target project, which are not able to be identified by any of our considered methods. The analysis results can help to reveal the performance bottlenecks of considered methods  in our study.

\noindent\textbf{Results.}
We first perform diversity analysis on identifying defective modules in the target projects between the HDP methods.
Then, we  perform diversity analysis on identifying defective modules in the target project between the unsupervised methods.
Finally, we perform  diversity analysis on identifying defective modules in the target project between the HDP methods and the unsupervised methods.
 Notice for the method UDP5, the choice of the optimal metric depends on the chosen performance measure. Therefore when the comparisons involve the unsupervised methods, the method UDP5 has two variants.
 In particular, UDP5-A denotes the choice of the optimal metric depends on the performance measure $\mathit{AUC}$, while UDP5-F denotes the choice of the optimal metric depends on the performance measure $\mathit{F1}$.

First,we analyze the diversity of prediction results on defective modules between the HDP methods. The final results can be found in Table~\ref{RQ3diversity1}. In this table, we can find the prediction diversity on defective modules exists in half of  HDP combinations except for HDP2 $vs.$ HDP4. In the best case, when HDP4 $vs.$ HDP5, prediction diversity on defective modules exists in 73.8\% of HDP combinations. When analyzing from the dataset group perspective, we can find the prediction diversity on
defective modules exists in half of CPDP pairs. In the best case, prediction diversity on defective modules
exists in 80\% of HDP combinations in AEEEM dataset group.

\begin{table*}[htbp]
  \centering
  \caption{Diversity Analysis on Defective Modules between HDP Methods}
    \begin{tabular}{ccccccc}\hline
    Method Comparison & Relink & AEEEM & SoftLab & PROMISE & NASA &  Summary  \\\hline
    HDP2 $vs.$ HDP3 & 70/93 & 90/145 & 68/145 & 96/240 & 174/339 & 498/962 \\
    HDP2 $vs.$ HDP4 & 29/93 & 41/145 & 13/145 & 51/240 & 103/339 & 237/962 \\
    HDP2 $vs.$ HDP5 & 73/93 & 116/145 & 80/145 & 169/240 & 252/339 & 690/962 \\
    HDP2 $vs.$ HDP1 & 41/63 & 67/91 & 65/136 & 96/154 & 116/179 & 385/623 \\
    HDP3 $vs.$ HDP4 & 75/93 & 86/145 & 74/145 & 113/240 & 176/339 & 524/962 \\
    HDP3 $vs.$ HDP5 & 72/93 & 113/145 & 76/145 & 155/240 & 260/339 & 676/962 \\
    HDP3 $vs.$ HDP1 & 58/63 & 67/91 & 97/136 & 116/154 & 124/179 & 462/623 \\
    HDP4 $vs.$ HDP5 & 73/93 & 116/145 & 86/145 & 163/240 & 272/339 & 710/962 \\
    HDP4 $vs.$ HDP1 & 46/63 & 62/91 & 74/136 & 103/154 & 127/179 & 412/623 \\
    HDP5 $vs.$ HDP1 & 44/63 & 69/91 & 74/136 & 117/154 & 150/179 & 454/623 \\\hline
   Summary & 581/810 & 827/1234 & 707/1414 & 1179/2056 & 1754/2750 & 5048/8264 \\\hline
    \end{tabular}%
  \label{RQ3diversity1}%
\end{table*}%

Then we analyze the diversity between the predicted results of the unsupervised methods. The final results can be found in Table~\ref{RQ3diversity2}.
 In this table, we can find the prediction diversity on defective modules is less obvious than the comparison between the HDP methods.
In particular, The prediction diversity only
 exists in half of  HDP combinations  for UM3 $vs.$ UM5-F, UM3 $vs.$ UM5-A, UM1 $vs.$ UM5-F, UM1 $vs.$ UM5-A, UM3 $vs.$ UM4. In the best case, when UM3 $vs.$ UM5-A, prediction diversity on defective modules
exists in 67.6\% of HDP combinations. When analyzing from the dataset group perspective, we can find the prediction diversity on
defective modules exists in half of CPDP pairs only in AEEEM and NASA dataset groups. In the best case, prediction diversity on defective modules
exists in 61.3\% of HDP combinations in AEEEM dataset group.


\begin{table*}[htbp]
  \centering
  \caption{Diversity Analysis on Defective modules between Unsupervised  Methods}
    \begin{tabular}{ccccccc}\hline
   Method Comparison & Relink & AEEEM & SoftLab & PROMISE & NASA & Summary\\\hline
      UM1 $vs.$ UM2 & 1/3   & 2/5   & 0/5   & 3/10  & 3/11  & 9/34 \\
    UM1 $vs.$ UM3 & 2/3   & 1/5   & 0/5   & 3/10  & 8/11  & 14/34 \\
    UM1 $vs.$ UM4 & 1/3   & 2/5   & 0/5   & 6/10  & 4/11  & 13/34 \\
    UM1 $vs.$ UM5-A & 1/3   & 5/5   & 0/5   & 4/10  & 7/11  & 17/34 \\
    UM1 $vs.$ UM5-F & 1/3   & 5/5   & 0/5   & 7/10  & 6/11  & 19/34 \\
    UM2 $vs.$ UM3 & 2/3   & 3/5   & 1/5   & 3/10  & 6/11  & 15/34 \\
    UM2 $vs.$ UM4 & 0/3   & 2/5   & 0/5   & 3/10  & 4/11  & 9/34 \\
    UM2 $vs.$ UM5-A & 0/3   & 4/5   & 0/5   & 5/10  & 4/11  & 13/34 \\
    UM2 $vs.$ UM5-F & 0/3   & 4/5   & 0/5   & 5/10  & 5/11  & 14/34 \\
    UM3 $vs.$ UM4 & 2/3   & 4/5   & 0/5   & 2/10  & 9/11  & 17/34 \\
    UM3 $vs.$ UM5-A & 2/3   & 5/5   & 1/5   & 4/10  & 9/11  & 21/34 \\
    UM3 $vs.$ UM5-F & 2/3   & 5/5   & 2/5   & 5/10  & 9/11  & 23/34 \\
    UM4 $vs.$ UM5-A & 0/3   & 1/5   & 0/5   & 1/10  & 6/11  & 8/34 \\
    UM4 $vs.$ UM5-F & 0/3   & 3/5   & 0/5   & 2/10  & 6/11  & 11/34 \\
    UM5-A $vs.$ UM5-F & 0/3   & 0/5   & 0/5   & 1/10  & 0/11  & 1/34 \\\hline
    Summary & 14/45 & 46/75 & 4/75  & 54/150 & 86/165 & 204/510 \\\hline
    \end{tabular}%
  \label{RQ3diversity2}%
\end{table*}%

Finally, we analyze the diversity of prediction results between the HDP and the unsupervised methods. The experimental results can be found in Table~\ref{RQ3diversity3}.
In this table, we can find the prediction diversity on defective modules is more obvious than the comparison between the HDP methods or between the unsupervised methods.
In particular, we can find the prediction diversity on defective modules exists in half of HDP combinations in all the HDP methods and the unsupervised methods comparisons. In the best case, when HDP1 $vs.$ UM5-F, prediction diversity on defective modules
exists in 93.6\% of HDP combinations. When analyzing from the dataset group perspective, we can find the prediction diversity on
defective modules exists in half of CPDP pairs except for SOFTLAB dataset groups.
In the best case, prediction diversity on defective modules
exists in 96.9\% of HDP combinations in AEEEM dataset group.


\begin{table*}[htbp]
  \centering
  \caption{Diversity Analysis on Defective Modules between the HDP Methods and Unsupervised CPDP Methods}
    \begin{tabular}{ccccccc}\hline
    Method Comparison & Relink & AEEEM & SoftLab & PROMISE & NASA & Summary\\\hline
    HDP1 $vs.$ UM1 & 59/63 & 86/91 & 107/136 & 130/154 & 170/179 & 552/623 \\
    HDP1 $vs.$ UM2 & 59/63 & 87/91 & 107/136 & 144/154 & 172/179 & 569/623 \\
    HDP1 $vs.$ UM3 & 58/63 & 87/91 & 98/136 & 138/154 & 165/179 & 546/623 \\
    HDP1 $vs.$ UM4 & 59/63 & 88/91 & 101/136 & 143/154 & 167/179 & 558/623 \\
    HDP1 $vs.$ UM5-A & 60/63 & 87/91 & 110/136 & 148/154 & 173/179 & 578/623 \\
    HDP1 $vs.$ UM5-F & 60/63 & 87/91 & 111/136 & 151/154 & 174/179 & 583/623 \\
    HDP2 $vs.$ UM1 & 80/93 & 143/145 & 56/145 & 160/240 & 305/339 & 744/962 \\
    HDP2 $vs.$ UM2 & 80/93 & 143/145 & 56/145 & 188/240 & 314/339 & 781/962 \\
    HDP2 $vs.$ UM3 & 74/93 & 144/145 & 39/145 & 165/240 & 289/339 & 711/962 \\
    HDP2 $vs.$ UM4 & 82/93 & 143/145 & 50/145 & 195/240 & 311/339 & 781/962 \\
    HDP2 $vs.$ UM5-A & 82/93 & 143/145 & 69/145 & 199/240 & 316/339 & 809/962 \\
    HDP2 $vs.$ UM5-F & 83/93 & 143/145 & 84/145 & 223/240 & 322/339 & 855/962 \\
    HDP3 $vs.$ UM1 & 90/93 & 145/145 & 44/145 & 99/240 & 278/339 & 656/962 \\
    HDP3 $vs.$ UM2 & 87/93 & 145/145 & 17/145 & 139/240 & 278/339 & 666/962 \\
    HDP3 $vs.$ UM3 & 69/93 & 145/145 & 41/145 & 81/240 & 274/339 & 610/962 \\
    HDP3 $vs.$ UM4 & 87/93 & 145/145 & 44/145 & 130/240 & 281/339 & 687/962 \\
    HDP3 $vs.$ UM5-A & 76/93 & 145/145 & 17/145 & 171/240 & 308/339 & 717/962 \\
    HDP3 $vs.$ UM5-F & 76/93 & 145/145 & 45/145 & 195/240 & 336/339 & 797/962 \\
    HDP4 $vs.$ UM1 & 66/93 & 145/145 & 55/145 & 161/240 & 301/339 & 728/962 \\
    HDP4 $vs.$ UM2 & 73/93 & 144/145 & 57/145 & 186/240 & 304/339 & 764/962 \\
    HDP4 $vs.$ UM3 & 60/93 & 145/145 & 28/145 & 167/240 & 284/339 & 684/962 \\
    HDP4 $vs.$ UM4 & 73/93 & 144/145 & 45/145 & 193/240 & 303/339 & 758/962 \\
    HDP4 $vs.$ UM5-A & 73/93 & 144/145 & 63/145 & 192/240 & 313/339 & 785/962 \\
    HDP4 $vs.$ UM5-F & 73/93 & 144/145 & 71/145 & 214/240 & 317/339 & 819/962 \\
    HDP5 $vs.$ UM1 & 79/93 & 127/145 & 89/145 & 161/240 & 296/339 & 752/962 \\
    HDP5 $vs.$ UM2 & 82/93 & 129/145 & 89/145 & 156/240 & 303/339 & 759/962 \\
    HDP5 $vs.$ UM3 & 77/93 & 123/145 & 85/145 & 156/240 & 280/339 & 721/962 \\
    HDP5 $vs.$ UM4 & 81/93 & 133/145 & 84/145 & 145/240 & 306/339 & 749/962 \\
    HDP5 $vs.$ UM5-A & 74/93 & 136/145 & 83/145 & 160/240 & 304/339 & 757/962 \\
    HDP5 $vs.$ UM5-F & 74/93 & 137/145 & 88/145 & 159/240 & 299/339 & 757/962 \\\hline
    Summary & 2206/2610 & 3902/4026 & 2033/4296 & 4849/6684 & 8243/9210 & 21233/26826 \\\hline
    \end{tabular}%
 \label{RQ3diversity3}%
\end{table*}%

In summary, the diversity of prediction results between HDP and unsupervised methods is more obvious. Therefore ensemble learning~\cite{zhou2012ensemble} is a potential way to further improve the performance of existing HDP methods.

Moreover, we analyze the defective modules in the project which can only be identified by one of these two kinds of methods, and the defect modules that cannot be identified by both methods. Due to the length limitation of this article, we only list the results when the projects in ReLink  group is set as the target project. The final results can be found in Table~\ref{RQ3}. In this table, since in some HDP
combinations, HDP1 cannot success (e.g., ant-1.3$\Rightarrow$Zxing), we only consider the HDP combinations HDP1 can success. ``== 0 by HDP" column and ``==0 by UM" column represent the number of defective modules which cannot be identified by these two methods, respectively. The ``proportion'' column represents the proportion of
the number of defective modules that cannot be identified by the method to the number of defective modules in the target project, The ``=0 by ALL'' column represents  the number of defective modules that cannot be identified by both two kinds of methods. For example, in this Table, when ant-1.3 is set to the source project and Apache is set to the target project, 96 defective modules in Apache cannot be identified by any HDP method, 71 defective modules cannot be identified by any unsupervised method, and moreover 55  defective modules cannot be identified by both HDP and unsupervised methods, which account for 19.48\%, 36.60\% and 28.35\% of the total number of Apache defective modules respectively. The results when the projects of other dataset groups can be found in  our project's website.
Based on the above results, we can find that there exist a certain number of  defective modules, which cannot be correctly predicted by either type of methods. Therefore we can conclude  the  HDP methods and the unsupervised methods still have some limitations in the ability to identify defective modules.


\begin{table*}[htbp]
  \centering
  \caption{Defective Modules Detection Ability for Different HDP Methods and Unsupervised Methods When the Projects in ReLink Group are set to the Target Projects}
     \begin{tabular}{lcccccc}\hline
    Source $\Rightarrow$ Target & \multicolumn{1}{l}{=0 by HDP} & \multicolumn{1}{l}{Proportion} & \multicolumn{1}{l}{=0 by UM} & \multicolumn{1}{l}{Proportion} & \multicolumn{1}{l}{=0 by ALL} & \multicolumn{1}{l}{Proportion} \\\hline
    ant-1.3$\Rightarrow$Apache & 96    & 49.48\% & 71    & 36.60\% & 55    & 28.35\% \\
    ant-1.3$\Rightarrow$Safe & 5     & 8.93\% & 23    & 41.07\% & 4     & 7.14\% \\
    ar1$\Rightarrow$Safe & 4     & 7.14\% & 23    & 41.07\% & 2     & 3.57\% \\
    ar1$\Rightarrow$Zxing & 93    & 23.31\% & 130   & 32.58\% & 23    & 5.76\% \\
    ar3$\Rightarrow$Apache & 5     & 2.58\% & 71    & 36.60\% & 0     & 0.00\% \\
    ar3$\Rightarrow$Safe & 0     & 0.00\% & 23    & 41.07\% & 0     & 0.00\% \\
    ar3$\Rightarrow$Zxing & 14    & 3.51\% & 130   & 32.58\% & 0     & 0.00\% \\
    ar4$\Rightarrow$Apache & 34    & 17.53\% & 71    & 36.60\% & 31    & 15.98\% \\
    ar4$\Rightarrow$Safe & 7     & 12.50\% & 23    & 41.07\% & 7     & 12.50\% \\
    ar4$\Rightarrow$Zxing & 260   & 65.16\% & 130   & 32.58\% & 112   & 28.07\% \\
    ar5$\Rightarrow$Apache & 90    & 46.39\% & 71    & 36.60\% & 50    & 25.77\% \\
    ar5$\Rightarrow$Safe & 3     & 5.36\% & 23    & 41.07\% & 3     & 5.36\% \\
    ar5$\Rightarrow$Zxing & 42    & 10.53\% & 130   & 32.58\% & 21    & 5.26\% \\
    ar6$\Rightarrow$Safe & 3     & 5.36\% & 23    & 41.07\% & 2     & 3.57\% \\
    ar6$\Rightarrow$Zxing & 192   & 48.12\% & 130   & 32.58\% & 68    & 17.04\% \\
    arc$\Rightarrow$Apache & 95    & 48.97\% & 71    & 36.60\% & 45    & 23.20\% \\
    arc$\Rightarrow$Safe & 3     & 5.36\% & 23    & 41.07\% & 2     & 3.57\% \\
    arc$\Rightarrow$Zxing & 285   & 71.43\% & 130   & 32.58\% & 107   & 26.82\% \\
    camel-1.0$\Rightarrow$Apache & 35    & 18.04\% & 71    & 36.60\% & 32    & 16.49\% \\
    camel-1.0$\Rightarrow$Safe & 0     & 0.00\% & 23    & 41.07\% & 0     & 0.00\% \\
    camel-1.0$\Rightarrow$Zxing & 271   & 67.92\% & 130   & 32.58\% & 120   & 30.08\% \\
    CM1$\Rightarrow$Apache & 8     & 4.12\% & 71    & 36.60\% & 6     & 3.09\% \\
    CM1$\Rightarrow$Safe & 0     & 0.00\% & 23    & 41.07\% & 0     & 0.00\% \\
    EQ$\Rightarrow$Apache & 6     & 3.09\% & 71    & 36.60\% & 6     & 3.09\% \\
    EQ$\Rightarrow$Safe & 1     & 1.79\% & 23    & 41.07\% & 1     & 1.79\% \\
    EQ$\Rightarrow$Zxing & 1     & 0.25\% & 130   & 32.58\% & 1     & 0.25\% \\
    JDT$\Rightarrow$Apache & 7     & 3.61\% & 71    & 36.60\% & 7     & 3.61\% \\
    JDT$\Rightarrow$Safe & 1     & 1.79\% & 23    & 41.07\% & 1     & 1.79\% \\
    JM1$\Rightarrow$Apache & 38    & 19.59\% & 71    & 36.60\% & 36    & 18.56\% \\
    JM1$\Rightarrow$Safe & 3     & 5.36\% & 23    & 41.07\% & 3     & 5.36\% \\
    KC3$\Rightarrow$Apache & 55    & 28.35\% & 71    & 36.60\% & 39    & 20.10\% \\
    KC3$\Rightarrow$Safe & 6     & 10.71\% & 23    & 41.07\% & 6     & 10.71\% \\
    KC3$\Rightarrow$Zxing & 281   & 70.43\% & 130   & 32.58\% & 112   & 28.07\% \\
    LC$\Rightarrow$Safe & 2     & 3.57\% & 23    & 41.07\% & 2     & 3.57\% \\
    LC$\Rightarrow$Zxing & 1     & 0.25\% & 130   & 32.58\% & 1     & 0.25\% \\
    MC1$\Rightarrow$Safe & 6     & 10.71\% & 23    & 41.07\% & 5     & 8.93\% \\
    MC2$\Rightarrow$Apache & 34    & 17.53\% & 71    & 36.60\% & 31    & 15.98\% \\
    MC2$\Rightarrow$Safe & 7     & 12.50\% & 23    & 41.07\% & 7     & 12.50\% \\
    MC2$\Rightarrow$Zxing & 173   & 43.36\% & 130   & 32.58\% & 63    & 15.79\% \\
    ML$\Rightarrow$Safe & 2     & 3.57\% & 23    & 41.07\% & 2     & 3.57\% \\
    ML$\Rightarrow$Zxing & 35    & 8.77\% & 130   & 32.58\% & 5     & 1.25\% \\
    MW1$\Rightarrow$Apache & 26    & 13.40\% & 71    & 36.60\% & 11    & 5.67\% \\
    MW1$\Rightarrow$Safe & 0     & 0.00\% & 23    & 41.07\% & 0     & 0.00\% \\
    MW1$\Rightarrow$Zxing & 2     & 0.50\% & 130   & 32.58\% & 0     & 0.00\% \\
    PC1$\Rightarrow$Safe & 0     & 0.00\% & 23    & 41.07\% & 0     & 0.00\% \\
    PC2$\Rightarrow$Safe & 0     & 0.00\% & 23    & 41.07\% & 0     & 0.00\% \\
    PC3$\Rightarrow$Safe & 0     & 0.00\% & 23    & 41.07\% & 0     & 0.00\% \\
    PC4$\Rightarrow$Safe & 1     & 1.79\% & 23    & 41.07\% & 1     & 1.79\% \\
    PC5$\Rightarrow$Safe & 3     & 5.36\% & 23    & 41.07\% & 3     & 5.36\% \\
    PDE$\Rightarrow$Safe & 1     & 1.79\% & 23    & 41.07\% & 1     & 1.79\% \\
    poi-1.5$\Rightarrow$Safe & 4     & 7.14\% & 23    & 41.07\% & 4     & 7.14\% \\
    skarbonka$\Rightarrow$Apache & 81    & 41.75\% & 71    & 36.60\% & 36    & 18.56\% \\
    skarbonka$\Rightarrow$Safe & 3     & 5.36\% & 23    & 41.07\% & 3     & 5.36\% \\
    skarbonka$\Rightarrow$Zxing & 56    & 14.04\% & 130   & 32.58\% & 10    & 2.51\% \\
    tomcat$\Rightarrow$Apache & 19    & 9.79\% & 71    & 36.60\% & 19    & 9.79\% \\
    tomcat$\Rightarrow$Safe & 5     & 8.93\% & 23    & 41.07\% & 5     & 8.93\% \\
    tomcat$\Rightarrow$Zxing & 192   & 48.12\% & 130   & 32.58\% & 110   & 27.57\% \\
    velocity-1.4$\Rightarrow$Safe & 4     & 7.14\% & 23    & 41.07\% & 0     & 0.00\% \\
    velocity-1.4$\Rightarrow$Zxing & 17    & 4.26\% & 130   & 32.58\% & 0     & 0.00\% \\
    xalan-2.4$\Rightarrow$Apache & 79    & 40.72\% & 71    & 36.60\% & 53    & 27.32\% \\
    xalan-2.4$\Rightarrow$Safe & 4     & 7.14\% & 23    & 41.07\% & 4     & 7.14\% \\
    xalan-2.4$\Rightarrow$Zxing & 293   & 73.43\% & 130   & 32.58\% & 112   & 28.07\% \\
    xerces-1.2$\Rightarrow$Safe & 11    & 19.64\% & 23    & 41.07\% & 9     & 16.07\% \\\hline
    \end{tabular}%
  \label{RQ3}%
\end{table*}%

\subsection{Results Analysis for RQ4}

\noindent\textbf{Motivation.}
In this RQ, we want to explore and verify the feasibility of the state-of-the-art HDP and unsupervised methods in our study by the means of computing satisfactory ratio of different kinds of HDP methods or unsupervised methods on a specified data set.

\noindent\textbf{Approach.}
The satisfactory criteria used by our study are mainly motivated by the previous study for CPDP. In particular, we use \textbf{SC1} to denote the criterion suggested by Zimmermann et al.~\cite{zimmermann:FSE2009} (i.e., $precision$ value and $recall$ value are both larger than 75\%) and \textbf{SC2} to denote the criterion suggested by He et al.~\cite{he2012investigation} (i.e., $recall$ value is larger than 70\% and $precision$ value is larger than 50\%). Both criteria are suggested in terms of two performance measures $precision$ and $recall$. Obviously, the first criterion is more rigorous than the second one.

\noindent\textbf{Results.}
The final results can be found in Table~\ref{RQ4}.
Here we compute
the satisfactory ratio of different  methods when the projects in a specific group are set to the target project when considering SC1 or SC2.
Taking ReLink dataset as an example, there are 31 (=5+10+11+5)  HDP combinations when considering the HDP5 method, if this method can achieve satisfactory
performance in 3 pairs, then the satisfactory ratio is 9.68\% (=3/31). Notice, the number of considered HDP combinations may less than 31 when analyzing the method HDP1, since HDP1 cannot succeed  in some of HDP combinations.
When considering SC1, all the HDP methods cannot achieve satisfactory performance when the projects of PROMISE, NASA, AEEEM are set to the target projects.
When considering SC2, all the HDP methods cannot achieve satisfactory performance when the projects of PROMISE and AEEEM are set to the target projects.
 Therefore, we conclude that the performance of the HDP methods (even the unsupervised methods) is still unsatisfactory. Improving the performance of the methods in this field is still a challenging task in the field of HDP.


\begin{table*}[htbp]
  \centering
  \caption{The Ratio of the HDP Methods and the Unsupervised Methods with Satisfactory Performance on the Five Groups of Datasets with Different Satisfactory Criteria}
    \begin{tabular}{lcccccccccc}\hline
    \multicolumn{1}{c}{\multirow{2}[0]{*}{Methods}} & \multicolumn{2}{c}{PROMISE} & \multicolumn{2}{c}{NASA} & \multicolumn{2}{c}{SoftLab} & \multicolumn{2}{c}{ReLink} & \multicolumn{2}{c}{AEEEM} \\\cline{2-11}
          & \multicolumn{1}{l}{SC1} & \multicolumn{1}{l}{SC2} & \multicolumn{1}{l}{SC1} & \multicolumn{1}{l}{SC2} & \multicolumn{1}{l}{SC1} & \multicolumn{1}{l}{SC2} & \multicolumn{1}{l}{SC1} & \multicolumn{1}{l}{SC2} & \multicolumn{1}{l}{SC1} & \multicolumn{1}{l}{SC2} \\\hline
    HDP1 & 0.00\% & 0.00\% & 0.00\% & 0.00\% & 0.74\% & 3.68\% & 0.00\% & 0.00\% & 0.00\% & 0.00\% \\
    HDP2 & 0.00\% & 5.38\% & 0.00\% & 0.00\% & 0.00\% & 0.69\% & 0.00\% & 0.00\% & 0.00\% & 0.00\% \\
    HDP3 & 0.00\% & 0.00\% & 0.00\% & 0.00\% & 0.00\% & 0.00\% & 0.00\% & 0.00\% & 0.00\% & 0.00\% \\
    HDP4  & 0.00\% & 9.68\% & 0.00\% & 0.00\% & 0.00\% & 0.00\% & 0.42\% & 0.42\% & 0.00\% & 0.00\% \\
    HDP5 & 0.00\% & 20.43\% & 0.00\% & 0.00\% & 0.00\% & 2.76\% & 0.42\% & 3.75\% & 0.00\% & 0.00\% \\
    UM1   & 0.00\% & 66.67\% & 0.00\% & 0.00\% & 0.00\% & 0.00\% & 0.00\% & 0.00\% & 0.00\% & 0.00\% \\
    UM2   & 0.00\% & 66.67\% & 0.00\% & 20.00\% & 0.00\% & 0.00\% & 0.00\% & 0.00\% & 0.00\% & 0.00\% \\
    UM3   & 0.00\% & 0.00\% & 0.00\% & 0.00\% & 0.00\% & 20.00\% & 0.00\% & 10.00\% & 0.00\% & 0.00\% \\
    UM4   & 33.33\% & 100.00\% & 0.00\% & 20.00\% & 0.00\% & 0.00\% & 0.00\% & 0.00\% & 0.00\% & 0.00\% \\
    UM5   & 0.00\% & 66.67\% & 0.00\% & 20.00\% & 0.00\% & 0.00\% & 0.00\% & 0.00\% & 0.00\% & 9.09\% \\\hline
    \end{tabular}%
  \label{RQ4}%
\end{table*}%

\section{Threats to Validity}

In this section, we mainly discuss the potential threats to validity in our empirical studies.

Threats to internal validity are mainly concerned with the uncontrolled internal factors that might have influence
on the experimental results.
To reduce this threat, For some of HDP methods and unsupervised methods (such as HDP1 to HDP4, UDP1 to UDP2), we utilize implementation shared by previous studies~\cite{Nam:ASE2015,li2018cost,li2017heterogeneous,LiASEJ2019}.  For some of methods (such as HDP5, UDP3 to UDP5), we implement their proposed methods and these methods have similar performance reported in related studies. For each considered method, we use the experimental setting suggested by corresponding study to guarantee the fairness of our empirical studies. Moreover, we also use simple example to examine the implementation correctness of all the effort-aware performance measures and non-effort-aware performance measure.

Threats to external validity are about whether the observed experimental results can be generalized to other
subjects.
To guarantee the representative of our empirical subjects, we chose data sets
which have been widely used in previous HDP research studies\cite{Nam:ASE2015,li2018cost,li2017heterogeneous,LiASEJ2019}. Moreover,  we choose five HDP methods and five unsupervised CPDP methods recently proposed in five years. Therefore, we can guarantee  the selected  methods can reflect the state of the art of HDP.

Threats to conclusion validity are mainly concerned with inappropriate use of statistical techniques.
In this article, We use BH corrected $p$-Value and Cliff¡¯s $\delta$ to examine whether the performance difference between two considered methods are statistically significant. Moreover, we use Scott-Knott test  to examine whether some methods outperform other methods and generate a global ranking of these considered methods.

Threats to construct validity are about whether the performance measures used in the empirical studies reflect the
real-world situation.
In this article, we not only investigate the non-effort-aware performance measures, but also consider the effort-aware performance measures. Therefore, we can have a holistic look at the performance comparison between the HDP methods and unsupervised methods.

\section{Conclusion and Future Work}
\label{chap6}

In this article, we perform a replication study to have a holistic look in state-of-the-art HDP method. In particular, we compare existing five HDP methods with state-of-the-art five unsupervised methods. Final results  show the HDP methods do not perform significantly better than the  unsupervised methods whether in terms of effort-aware performance measures or non-effort-aware performance measures. Therefore, there is still a long way for HDP to go and in the future and we  also suggest future studies on HDP should consider unsupervised methods (especially simple method UDP4) as baselines when evaluating new designed HDP methods.

In the future, we plan to extend our research in two ways. First, we want to investigate the generalization of our empirical results by considering more software projects. Second this study shows there is a
long way to go for HDP,  therefore more
effective HDP methods  should be designed in the future.




\section*{Acknowledgements}
Xiang Chen and Yanzhou Mu have contributed equally for this work and they are co-first authors.
Chao Ni is the corresponding author.
This work is
supported in part by National Natural Science Foundation of China
(Grant nos. 61702041, 	61602267 and 61202006), and Open Project of State
Key Laboratory for Novel Software Technology at Nanjing University
(Grant no. KFKT2019B14), The Nantong Application Research Plan (Grant no. JC2018134) and
China Scholarship Council (CSC) funding the visit of Chao Ni (No.201806190172) to Monash University.
More details of empirical results can be found in the website of our project\footnote{The website will be given when the manuscript is accepted.}.

%
%


\bibliographystyle{spmpsci}
\bibliography{IEEEabrv,Bibliography}

\vfill

\end{document}